\documentclass[aps,preprint,nofootinbib,longbibliography,a4paper,preprintnumbers,superscriptaddress,eqsecnum]{revtex4-1}
\usepackage{color}

\newcommand{\nb}[1]{\color{blue}}

\newcommand{\hl}[1]{\color{magenta}}

\usepackage[
	  pagebackref=false,
	  colorlinks=true,
      linkcolor=blue,
      urlcolor=blue,
      filecolor=black,
      citecolor=red,
      pdfstartview=FitV,
      pdftitle={},
        pdfauthor={},
        pdfsubject={},
        pdfkeywords={},
        pdfpagemode=None,
        bookmarksopen=true
      ]{hyperref}

\usepackage[T1]{fontenc}
\usepackage[utf8]{inputenc}

\usepackage[normalem]{ulem}
\usepackage{amsmath}
\usepackage{enumerate}
\usepackage{amsfonts}
\usepackage{epsfig}
\usepackage{mathbbol}

\usepackage{graphicx} 
\usepackage{amsthm}
\usepackage{amssymb} 
\usepackage{dsfont}
\usepackage{yfonts}
\usepackage{hyperref}
\usepackage{array,xcolor,graphicx}
\usepackage{mathtools}

\newcommand\p{\ensuremath{\partial}}

\newcommand{\be}{\begin{equation}}
\newcommand{\ee}{\end{equation}}
\newcommand{\bea}{\begin{eqnarray}}
\newcommand{\eea}{\end{eqnarray}}
\newcommand{\bega}{\begin{gather}}
\newcommand{\eega}{\end{gather}}

\newcommand{\bi}{\begin{itemize}}
\newcommand{\ei}{\end{itemize}}
\newcommand{\ben}{\begin{enumerate}}
\newcommand{\een}{\end{enumerate}}
\newcommand{\bca}{\begin{cases}}
\newcommand{\eca}{\end{cases}}
\newcommand{\bln}{\begin{align}}
\newcommand{\eln}{\end{align}}
\newcommand{\bst}{\begin{split}}
\newcommand{\est}{\end{split}}
\def\ie{\begin{equation}\begin{aligned}}
\def\fe{\end{aligned}\end{equation}}
\newcommand{\bma}{\le(\begin{matrix}}
\newcommand{\ema}{\end{matrix}\ri)}

\newcommand\al{{\alpha}}
\def\b{{\beta}}
\newcommand\ep{\epsilon}

\newcommand\lam{\lambda}

\newcommand\om{\omega}

\newcommand\de{{\ensuremath{{\delta}}}}

\newcommand\ov{\over}

\def\le{\left}
\def\ri{\right}

\newcommand\sL{{\ensuremath{{\mathcal L}}}}

\newcommand\vx{{\vec x}}

\newcommand\neww[1]{{\color{red} #1}}

\begin{document}

\title{\large Many-body chaos and energy dynamics in holography}

\author{Mike Blake} 
\affiliation{Center for Theoretical Physics, \\
Massachusetts Institute of Technology,
Cambridge, MA 02139, USA }

\author{Richard A. Davison}
\affiliation{Department of Physics, \\
 Harvard University, Cambridge, MA 02138, USA 
 \vspace{0.5cm}
}
 
\author{Sa\v{s}o Grozdanov}
\affiliation{Center for Theoretical Physics, \\
Massachusetts Institute of Technology,
Cambridge, MA 02139, USA }

\author{Hong Liu}
\affiliation{Center for Theoretical Physics, \\
Massachusetts Institute of Technology,
Cambridge, MA 02139, USA }

\preprint{MIT-CTP/5046}

\begin{abstract}
\vspace{0.5cm}
\noindent   Recent developments have indicated that in addition to out-of-time ordered correlation functions (OTOCs), quantum chaos also has a sharp manifestation in the thermal energy density two-point functions, at least for maximally chaotic systems. The manifestation, referred to as pole-skipping, concerns the analytic behaviour of energy density two-point functions around a special point $\omega = i \lambda$, $k = i \lambda/v_B$ in the complex frequency and momentum plane. Here $\lambda$ and $v_B$ are the Lyapunov exponent and butterfly velocity characterising quantum chaos. In this paper we provide an argument that the phenomenon of pole-skipping is universal for general finite temperature systems dual to Einstein gravity coupled to matter. In doing so we uncover a surprising universal feature of the linearised Einstein equations around a static black hole geometry. We also study analytically a holographic axion model where all of the features of our general argument as well as the pole-skipping phenomenon can be verified in detail. 
\end{abstract}

%\today

\maketitle

\begingroup
\hypersetup{linkcolor=black}
\tableofcontents
\endgroup

\section{Introduction}

Over the last few years there has been exciting progress in characterizing chaotic behavior in quantum many-body systems
using out-of-time ordered correlation functions (OTOCs) \cite{1969JETP...28.1200L,Shenker:2013pqa,Roberts:2014isa,Shenker:2014cwa,Maldacena:2015waa,kitaev,Kitaev:2017awl,Polchinski:2015cea,Polchinski:2016xgd,Jensen:2016pah,Maldacena:2016hyu, Maldacena:2016upp, Gu:2016oyy, Davison:2016ngz,Patel:2016wdy,Aleiner:2016eni,Nahum:2017yvy, Khemani:2018sdn,Kukuljan:2017xag} 
For instance, for a large class of systems,  it has been observed that\footnote{We have normalized $V$ and $W$ so that at $t=0$, the correlator is $1$. Some variances of~\eqref{otoc} have also been observed. For example,  $\lam$ can be zero in some systems and 
different spatial profiles have also been seen. See for example~\cite{Khemani:2018sdn} for a survey.} 
\be
\langle V(t,\vx) W(0) V(t,\vx) W(0) \rangle_{\beta_0} =  1 - \ep \, e^{\lambda (  t - |\vx|/v_B ) } + \cdots,
\label{otoc}
\ee
where $V$ and $W$ are generic few-body operators, $\beta_0 = 1/T$ is the inverse temperature, and $\ep$ is a small parameter inversely proportional to the number of degrees of freedom. The exponential growth of~\eqref{otoc} is reminiscent of the diverging trajectories of two initially infinitesimally separated particles in classical chaotic systems. Thus 
$\lambda$ is often referred to as the quantum Lyapunov exponent, and $v_B$,  which describes the speed at which the growth propagates in space, as the butterfly velocity. For later purposes, let us note that the equation~\eqref{otoc} has the form of a plane wave
\begin{align}
\langle V(t,\vx) W(0) V(t,\vx) W(0) \rangle_{\beta_0} \sim   1 - \ep e^{- i \om t + i k |\vx|} + \cdots ,
\end{align}
with purely imaginary values of both frequency $\omega$ and momentum $k$:
\begin{align}
\omega = i \lambda \,, \qquad k = i k_0 \,, \qquad  k_0 = \frac{\lambda }{v_B} \, .
\label{location}
\end{align}

The quantum nature of~\eqref{otoc} is  highlighted by the fact that $\lam$ has an upper bound $\lambda \leq \lam_{\rm max} = \frac{2 \pi k_B}{\hbar \beta_0}$~\cite{Maldacena:2015waa} (henceforth $\hbar = k_B = 1$), 
which is saturated by a variety of systems including holographic theories,
 two-dimensional conformal field theories (CFTs) in the large central charge limit, and strongly coupled SYK models. 
Below we will refer to systems which saturate the bound as maximally chaotic systems. 

At the classical level, chaos is believed to provide the microscopic dynamical mechanism for macroscopic phenomena such as transport and thermalisation (for example see~\cite{ott_2002,gaspard_1998}). It remains an outstanding open question 
whether chaos plays a similar role for quantum many-body systems. It is thus a welcome development that
there have been various tantalising hints indicating that~\eqref{otoc} is inextricably connected to transport and hydrodynamic behavior~\cite{Blake:2016wvh,Blake:2016sud,Blake:2016jnn,Gu:2016oyy,Patel:2016wdy,Davison:2016ngz,Blake:2017qgd,Grozdanov:2017ajz,Blake:2017ris,Grozdanov:2018atb,Lucas:2017ibu,Hartman:2017hhp,Davison:2018ofp}. 

In particular, two recent developments showed that in addition to~\eqref{otoc}, quantum chaos also has a sharp manifestation in the thermal energy density two-point functions~\cite{Grozdanov:2017ajz,Blake:2017ris}. The manifestation, referred to as pole-skipping in~\cite{Blake:2017ris}, was first observed numerically in a holographic system in~\cite{Grozdanov:2017ajz}, and then derived as a general prediction of an effective field theory (EFT) for chaos~\cite{Blake:2017ris}. More explicitly, it was proposed in~\cite{Blake:2017ris} that the chaotic behavior~\eqref{otoc} can be captured by the propagation of an effective chaos mode, which, at least for maximally chaotic systems, coincides precisely with the 
hydrodynamic mode for energy conservation. The phenomenon of pole-skipping is a direct consequence of the same mode playing  the dual role of capturing both energy conservation and chaos.  The EFT description of chaos also provides a simple explanation for 
connections between~\eqref{otoc} and energy diffusion observed in previous literature \cite{Blake:2016wvh,Blake:2016sud,Blake:2016jnn,Blake:2017qgd,Blake:2017ris,Gu:2016oyy,Davison:2016ngz,Grozdanov:2017ajz}; they correspond to the behaviour of a single effective mode, albeit at different scales: the OTOC \eqref{otoc} at the scale given by~\eqref{location}, and transport at scales of $\om, k \ll1/\b_0$. They are related by an $O(1)$ extrapolation. 

Pole-skipping can be checked to hold for the SYK chains studied in~\cite{Gu:2016oyy}. And 
it has also been verified in two-dimensional CFTs in the large central charge limit~\cite{moshe}. 

The purpose of this paper is to provide further support for the generality of the manifestation of quantum chaos in energy density correlation functions. We show that the pole-skipping phenomenon is universal in general  holographic systems described by the Einstein gravity coupled to matter fields, thereby generalising the analysis of \cite{Grozdanov:2017ajz} for the Schwarzschild black hole in AdS$_5$. We will provide a general analytic derivation of this phenomenon and elucidate its gravitational origin. We will also make a connection with the gravitational shock wave analysis of~\eqref{otoc}, thus directly establishing that the behaviour of an OTOC and pole-skipping have the same gravitational origin. In doing so we will uncover a surprising universal feature of the linearised Einstein's equations coupled to matter around a static geometry. A hint of this universal feature came from the analysis of~\cite{Grozdanov:2017ajz} which initiated the study of linearised gravitational perturbations at the special point~\eqref{location}. We emphasise that for a general gravity system there is no near-horizon $SL(2,R)$ symmetry (as there is in AdS$_2$ or the BTZ black hole). This makes it clear that pole-skipping is a phenomenon that occurs more generally than just in highly symmetric cases.

Before summarising the main contents of the paper, let us first review the phenomenon of pole-skipping in  the energy density two-point functions. The simplest example which exhibits this phenomenon is the SYK chain of \cite{Gu:2016oyy} for which the energy density retarded two-point function has the form  
\begin{align}
G^R_{T^{00} T^{00} }(\omega,k) = C {i \om \le( {\om^2 \ov \lam^2} + 1\ri) \ov - i \om + D_E k^2} \, ,  
 \label{ttsy}
\end{align}
where $C$ is some constant and $D_E$ is the energy diffusion constant. For this system the Lyapunov exponent $\lam = \lam_{\rm max}$ and $v_B = \sqrt{\lam_{\rm max} D_E}$. Equation~\eqref{ttsy} has a line of diffusion poles at $\om = - i D_E k^2$. Now, analytically continuing $\om$ and $k$ to imaginary values, we see that the pole line precisely passes through the special point~\eqref{location}. But at that point, the numerator of~\eqref{ttsy} is zero, so at the special point the would-be pole skips. It has been argued in~\cite{Blake:2017ris} that this is a generic phenomenon for maximally chaotic systems. More generally, writing 
\be
G^R_{T^{00} T^{00} }(\omega,k) = {b(\om, k) \ov a (\om, k)} \, ,
\ee
it follows from energy conservation that $G^R$ should exhibit hydrodynamic poles at small $\om, k$, i.e. $a(\om, k)$ should have a line of zeros at 
\be \label{poln}
\om = \om_h (k) = \bca c_s k + \cdots \cr - i D_E k^2 + \cdots   \eca
\ee
where in the second equality we have performed a small $k$ expansion with the upper (lower) line for a system with (without) momentum conservation. Now the statement of pole-skipping is: 
\ben 
\item  When analytically continued to imaginary $\om$ and $k$,~\eqref{poln} passes precisely through the special point~\eqref{location}; 

\item $b(\om,k)$ has a line of zeros which also passes through the special point~\eqref{location}. 
\een
The same phenomenon is also present in $G^R_{T^{00} T^{0i}}, G^R_{T^{0i} T^{0i}}$ as they are related by Ward identities. For the rest of paper we will simply focus on $G^R_{T^{00} T^{00}}$.

An immediate consequence of having zeros of $a(\om, k)$ and $b(\om,k)$ passing through the same point~\eqref{location} is that there, $G^R_{T^{00} T^{00}} = {0 \ov 0}$, and thus the energy density two-point does {\it not} have a well defined value at that specific $\omega$ and $k$! More explicitly, let us consider 
\be 
\om = i \lam + \de \om, \qquad k = i k_0 + \de k \,.
\ee
Then, as $\de \om, \de k \to 0$, we find 
\be \label{muk}
G^R_{T^{00} T^{00} }(\omega,k) = {\p_\om b (i \lam, i k_0) {\de \om \ov \de k} + \p_k b(i \lam, i k_0)  \ov 
 \p_\om a(i \lam, i k_0) {\de \om \ov \de k} + \p_k a(i \lam, i k_0) } \ .
 \ee
 In other words, the function becomes infinitely multiple-valued at~\eqref{location}, depending on the slope $ \de\om /\de k$ at which the point is approached. Thus if the gravity analysis is to reproduce the pole-skipping phenomenon, something special must be happening to the linearised Einstein's equations at~\eqref{location}. We will see this indeed to be the case. 

We now summarise the main results of the paper on holographic systems. 

In Section~\ref{sec:generalargument} we study the linearised Einstein's equations around a general black brane geometry in ingoing Eddington-Finkelstein coordinates, and demonstrate that precisely at~\eqref{location} these equations exhibit a remarkably universal behaviour near the horizon. Specifically, we will show that one component of the near-horizon Einstein's equations becomes trivial at the point~\eqref{location}, and hence one finds an extra ingoing mode near \eqref{location}. 
As a result, in the neighbourhood of~\eqref{location} we find a family of different ingoing modes that depend on the slope $\delta \omega/\delta k$ with which one approaches \eqref{location}, precisely mirroring the situation of~\eqref{muk}. By choosing this slope one can find a normalisable mode that passes through the location \eqref{location} and hence there must be a pole in the energy density two-point correlator passing through \eqref{location}. Likewise, by choosing a different slope one can also find a different ingoing solution corresponding to a line of zeroes passing through \eqref{location}.

To illustrate the general discussion of Section~\ref{sec:generalargument}, in Section~\ref{sec:greensfunction} we proceed to study this phenomenon in detail in a specific holographic model which describes a class of three-dimensional strongly coupled field theories with broken translational symmetries~\cite{Andrade:2013gsa}. More explicitly, the model has a continuous parameter $m$. For $m=0$, it reduces to a translationally invariant black brane in  AdS$_4$ describing a three-dimensional CFT at a finite temperature. For $m \neq 0$, spatial momenta are not conserved, with dimensionless parameter ${m \beta_0}$ controlling the strength of momentum dissipation. We choose this model as it exhibits rich transport behaviour as one varies $m$ (see Ref. \cite{Davison:2014lua}), thus allowing us to demonstrate explicitly the insensitivity of the pole-skipping phenomenon to the long distance transport properties of a system (as long as energy is conserved).  Yet the model is simple enough to allow us to calculate the energy density two-point functions analytically, which in turn enables us to exhibit all features of the general argument of Sec.~\ref{sec:generalargument} as well as the pole-skipping phenomenon 
in great detail. In Appendix~\ref{app:ads5} we show that the same analytic discussion can also be applied to the Schwarzschild black hole of AdS$_5$, dual to a four-dimensional CFT at a finite temperature. In Sec.~\ref{sec:discussion} we conclude with a discussion of future directions.

\section{Near-horizon Einstein's equations and pole-skipping} 

\label{sec:generalargument}

In this section, we will demonstrate a remarkably universal property of the linearised Einstein's equations coupled to general matter content. We will then use this property to argue that in general, the retarded energy density two-point correlator of holographic theories with Einstein gravity exhibits pole-skipping at the location of Eq. \eqref{location}. This property also enables us to make a connection with the gravitational shock wave analysis \cite{Shenker:2013pqa,Roberts:2014isa,Shenker:2014cwa} of~\eqref{otoc}, thus establishing that the behaviour of the OTOC and pole-skipping have the same gravitational origin. The discussion of this section is general, thus somewhat abstract. In the next section we will then examine a family of examples very explicitly.

\subsection{Setup} 

Let us start by considering Einstein's equations that arise from a bulk action of the form 
\begin{equation}
S=\int d^{d+2}x\sqrt{-g}\left(\mathcal{R}-2\Lambda+\mathcal{L}_M\right),
\label{eq:action}
\end{equation}
where $\Lambda=-d(d+1)/2L^2$, $\mathcal{L}_M$ is the matter Lagrangian and $L$ is the AdS radius that henceforth we set to 1. We will allow for a rather general matter Lagrangian $\mathcal{L}_M$, but will assume that the theory admits an equilibrium configuration corresponding to a homogeneous and isotropic black brane. We write the background metric as 
\begin{equation}
\label{eq:background}
ds^2 = -r^2 f(r) dt^2 + \frac{dr^2}{r^2 f(r)} + h(r)d\vec{x}^2\,,
\end{equation}
where $f(r)$ is the emblackening factor which vanishes at the horizon $f(r_0) = 0$. The Hawking temperature is given by $T = r_0^2 f'(r_0)/ (4\pi)$. 

Note that for a metric of this form the special frequency and momentum in \eqref{location} are extracted by constructing the Dray-`t Hooft shock wave describing a gravitational perturbation $\delta g_{VV} = c(x) \delta(V)$ where $V$ is the ingoing Kruskal-Szekeres coordinate.  In the presence of an infalling matter perturbation with energy $E$ thrown into the black hole at time $t = - t_{w}$, the $VV$ component of Einstein's equations leads to a decoupled equation for $c(x)$ \cite{Shenker:2013pqa,Roberts:2014isa,Shenker:2014cwa,Roberts:2016wdl,Blake:2016wvh}
\begin{equation}
\label{thooft}
(\nabla^2 - d \pi T h'(r_0)) c(x) \sim e^{2 \pi t_{w}/\beta_0} E \delta (x) \,,
\end{equation}
where $\beta_0 = 1/T$ is the inverse temperature. Solving this equation gives an exponential profile for $c(x)$ that leads to the form \eqref{otoc} with the Lyapunov exponent and butterfly velocity given by 
\begin{align}
\label{eq:location0}
\lambda = 2 \pi T \,, && k_0^2 = \frac{(2\pi T)^2}{v_B^2} = d \pi T h'(r_0) \,.
\end{align}

Here we wish to study the retarded energy density correlation function $G^{R}_{T^{00} T^{00}} (\omega, k)$ near \eqref{eq:location0}. This correlation function can be extracted from solving the linearised gravitational perturbation equations around \eqref{eq:background} subject to ingoing boundary conditions at the horizon. It is therefore convenient to introduce ingoing Eddington-Finkelstein (EF) coordinates $(v, r, x^i)$ with $x^1\equiv x$ and
\begin{align}
v = t + r_*  \,,&& \frac{dr_*}{dr} = \frac{1}{r^2 f(r)} \,,
\end{align}
in terms of which the background metric \eqref{eq:background} is 
\begin{equation}
\label{eq:backgroundef}
ds^2 = - r^2 f(r) dv^2 + 2 dv dr + h(r)d\vec{x}^2\,.
\end{equation}
To calculate the energy density correlation function we then need to study the perturbation equations of the metric $\delta g_{vv}(r,v,x) = \delta g_{vv}(r) e^{- i \omega v + i k x}$, together with all the other metric perturbations and matter fields that couple to this mode. The fields that couple in this channel are $\delta g_{vv}$, $\delta g_{rr}$, $\de g_{vx}$, $\delta g_{vr}$, $\delta g_{x^ix^i}$, $\delta g_{rx}$ and $\delta \varphi$,  where $\delta \varphi$ schematically represents any matter fields that couple to these perturbations.\footnote{This requires that $\delta T_{vx^i},\delta T_{rx^i},\delta T_{x^1x^i},\delta T_{x^jx^k}$ all vanish, with $i\ne1$ and $j\ne k\ne1$.} 

\subsection{Near-horizon expansion} 

The retarded Green's function of energy density $G^R_{T^{00}T^{00}}$ is governed by solutions of the gravitational and matter perturbation equations that are regular at the horizon in ingoing EF coordinates. It is therefore useful to expand these perturbations near the horizon as
\begin{eqnarray}
\label{eq:irexp}
\delta g_{\mu \nu}(r) &=& \delta g_{\mu \nu}^{(0)} + \delta g_{\mu \nu}^{(1)} (r - r_0) + \ldots, \nonumber \\ 
\delta \varphi(r) &=& \delta \varphi^{(0)} + \delta \varphi^{(1)} (r - r_0) + \ldots  \,.
\end{eqnarray}
After inserting the expansion \eqref{eq:irexp} into Einstein's equations one can construct the most general solution to the perturbation equations as an expansion about the horizon.

Since the special location \eqref{location} (with parameters there given by~\eqref{eq:location0}) depends only on the background metric at the horizon, then if there is to be any universal behaviour near this point it must arise from the near-horizon behaviour of these equations. To examine this we simply insert \eqref{eq:irexp} into the perturbations equations and evaluate them at the horizon, which gives a set of constraints that relate the near-horizon coefficients in \eqref{eq:irexp} to each other. The full details of these equations are quite complicated and are not necessary for our purposes. Thus, in order to illustrate the key point it is enough to proceed by dividing them into two classes: the $vv$ component of Einstein's equations $E_{vv} = 0$ and all other perturbation equations, which we schematically write as $X = 0$. 

As we approach the point \eqref{location} the perturbation equations $X = 0$ remain well defined and, together with the perturbation equations away from the horizon, impose non-trivial constraints that must be satisfied by the solution \eqref{eq:irexp}. In contrast, the behaviour of the $vv$ component of the Einstein equations is highly singular near \eqref{location}. In particular, evaluating this equation at the horizon we find that 
\begin{align}
\label{eq:equationwithmatter}
\le(- i \frac{d}{2}\omega h'(r_0) + k^2 \ri) \delta g_{vv}^{(0)} &- i (2 \pi T + i \omega) \left[  \omega \delta g_{x^ix^i}^{(0)} + 2 k \delta g_{vx}^{(0)} \right]  \nonumber  \\
&= -2 h(r_0) \bigg[T_{vr}(r_0)\delta g_{vv}^{(0)} - \delta T_{vv}(r_0) \bigg] \,,
\end{align}
where $T_{\mu \nu}(r_0)$ is the bulk stress-energy tensor of the background matter fields supporting the black brane \eqref{eq:background} and $\delta T_{\mu\nu }(r_0)$ describes the matter perturbations. A priori, \eqref{eq:equationwithmatter} therefore depends on the matter content of the theory, i.e. the precise form of $\sL_M$ in~\eqref{eq:action}. However, for a large class of black brane solutions that we examined, including AdS gravity coupled to scalar and gauge fields, we find that the {\it identity}
\begin{equation}
\label{eq:stresstensorcondition}
\bigg[T_{vr}(r_0)\delta g_{vv}^{(0)} - \delta T_{vv}(r_0) \bigg] = 0,
\end{equation}
holds automatically for any value of $\omega$ and $k$ as a consequence of regularity at the horizon. In particular, in Appendix~\ref{app:axiondilaton} we show that the identity \eqref{eq:stresstensorcondition} holds for the commonly studied Einstein-Maxwell-Dilaton-Axion gravity theories. As such, in all these theories the $vv$ component of Einstein's equations at the horizon takes a remarkably universal form that depends only on the metric perturbations
\begin{equation}
\label{perteqn0}
\le(- i \frac{d}{2}\omega h'(r_0) + k^2\ri) \delta g_{vv}^{(0)} - i (2 \pi T + i \omega) \bigg[ \omega \delta g_{x^ix^i}^{(0)}  + 2  k  \delta g_{vx}^{(0)} \bigg] = 0 \,.
\end{equation}

\subsection{Solutions at special point} 

From \eqref{perteqn0} we can now see that the location \eqref{location} in Fourier space is indeed very special. In particular, note that for general $\omega$ and $k$ equation \eqref{perteqn0} provides a non-trivial constraint that relates the parameters $\delta g_{vv}^{(0)}, \delta g_{vx}^{(0)}, \delta g_{x^ix^i}^{(0)}$ of the near-horizon solution to each other. However, when $\omega = 2 \pi T i$ then all other fields decouple at the horizon from $\delta g_{vv}^{(0)}$. The equation \eqref{perteqn0} then reduces to
\begin{equation}
\label{shock}
(d \pi T h'(r_0) + k^2) \delta g_{vv}^{(0)} = 0 \,,
\end{equation}
which remarkably has the same form as the equation that determines the spatial profile of the Dray-`t Hooft shock wave \eqref{thooft}. For generic $k$ this equation sets $\delta g_{vv}^{(0)} = 0$, however for $k = i k_0$ then this component of Einstein's equations is automatically satisfied. This means that precisely at the special point \eqref{location}, equation \eqref{perteqn0} is identically zero, and hence does not impose any constraint on the near-horizon expansion parameters $\delta g_{vv}^{(0)}$, $\delta g_{vx}^{(0)}$ or $\delta g_{x^ix^i}^{(0)}$. In other words, precisely at \eqref{location} there is one fewer equation to solve at the horizon than at a generic value of $\omega$ and $k$.

As a consequence we can deduce that at the location \eqref{location} there exists an extra linearly independent ingoing solution to Einstein's equations. For the specific cases of AdS$_4$ gravity or the axion model that we study in detail in Section~\ref{sec:greensfunction} we can check this by explicitly constructing the solutions \eqref{eq:irexp} by solving the equations of motion order-by-order in the expansion \eqref{eq:irexp}. For instance, in AdS$_4$ gravity then, after fixing radial gauge, we find that the most general ingoing solution to Einstein's equations is specified at generic $\omega$ and $k$ by 4 parameters in this near-horizon expansion. These can be chosen to be the parameters $\delta g_{vx}^{(0)}$, $\delta g_{xx}^{(0)}$, $\delta g_{yy}^{(0)}$ and $\delta g_{yy}^{(1)}$ (where $x^2\equiv y$). However, precisely at \eqref{location}, we find that there is an extra independent parameter, which can be chosen to be the metric component $\delta g_{vv}^{(0)}$ at the horizon. In total the most general solution to the equations of motion then has 5 independent parameters in the near-horizon expansion at \eqref{location}.

The consequences of having such an extra linearly independent solution at \eqref{location} are dramatic. With a slight risk of oversimplifying the problem, but in aid of conceptual clarity, let us use a single scalar field as an illustration of the implications this extra solution. Recall that a scalar field $\phi$ has the following expansion near the boundary $r \to \infty$:
\be 
\phi (r, \om, k) = A(\om, k) r^{-\al_1} + B (\om, k) r^{-\al_2}\,, \qquad \al_1 < \al_2 \,,
\ee
where the $A$-term ($B$-term) is the non-normalisable (normalisable) term and corresponds to the external source (expectation value). The dual retarded Green's function is then given by $G_R = B/A$ with $\phi$ obeying ingoing boundary conditions at the horizon \cite{Son:2002sd}. For generic $\om$ and $k$, at the horizon, $\phi$ has an ingoing and an outgoing mode. When we choose the ingoing mode, the ratio of $B$ over $A$ becomes completely fixed, and so does the retarded Green's function. Suppose now that at the special point~\eqref{location}, we have an extra ingoing mode at the horizon, which means that both independent solutions of $\phi$ are now allowed at the horizon. Then, there is no constraints on $A$ or $B$, and the retarded Green's function becomes infinitely multiple-valued, depending on one free parameter which determines the linear combination of two near-horizon solutions. In particular, we can always choose a combination so that there is only the normalisable term near the boundary (i.e. $A=0$), which then leads to a pole in $G_R$. Similarly, there exists another combination such that the normalisable term is absent (i.e. $B=0$), which then corresponds to a zero of $G_R$.

The present situation with metric perturbations is a bit more complicated, but the essence is the same: the extra freedom in $\de g_{vv}$ at the horizon should {generically} lead to one free parameter in $G^R_{T^{00} T^{00}}$ at the special point. We will shortly see that when we move slightly away from the special point, the free parameter can be taken to be the slope $\de \om / \de k$ approaching the point,  thereby precisely mirroring the situation of equation~\eqref{muk}.
In fact, after fixing all the bulk gauge freedom and solving the constraints, the sector {of metric perturbations} associated with $\de g_{vv}$ always reduces to a single scalar degree of freedom \cite{Kovtun:2005ev}. On the boundary theory side this can be understood as a result of Ward identities for $G^R_{T^{00} T^{00}}$,  $G^R_{T^{00} T^{0x}}$, $G^R_{T^{0x} T^{0x}}$, whereby all of these two-point functions are proportional to a single scalar function. We will see this explicitly in the example studied in the next section. 

To close this subsection, we note that the fact that at this special point the metric component $\delta g_{vv}^{(0)}$ can be tuned independently of the other fields at the horizon resonates with the analysis of \cite{Grozdanov:2017ajz} which found a special null ansatz solution to Einstein's equations at \eqref{location}. The solution constructed in \cite{Grozdanov:2017ajz} is indeed a special case of our general ingoing solutions corresponding to the case with $\delta g_{x^ix^i}^{(0)} = \delta g^{(0)}_{vx} = 0$ and $\delta g_{vv}^{(0)}$ non-zero. However, we emphasize that from the point of view of the present analysis the key feature of \eqref{location} is not that ingoing solutions with only $\delta g_{vv}^{(0)} \neq 0$ at the horizon exist at \eqref{location}, but rather that such a solution represents an additional independent solution to the equations of motion.

\subsection{Solutions near special point} 

As we have seen, \eqref{perteqn0} becomes trivial at \eqref{location} and leads to an extra ingoing solution precisely at this point. As a result the Green's function $G^R_{T^{00} T^{00}}$ is ill defined precisely at \eqref{location}, and is infinitely multivalued. To understand physically what this means for the retarded Green's function, we can consider tuning $\omega$ and $k$ slightly away from \eqref{location}. That is we consider Einstein's equations with $\omega = i \lambda + \epsilon\delta \omega$ and $k = i k_0 + \epsilon \delta k$, for $|\epsilon| \ll 1$. Then at leading order in $\epsilon$ the solution must satisfy the smooth equations $X=0$ together with an extra non-trivial constraint arising from \eqref{perteqn0} 
\begin{equation}
\bigg(-  \frac{d}{2}\delta \omega h'(r_0) + 2  k_0 \delta k \bigg) \delta g_{vv}^{(0)} +  \delta \omega \bigg[ 2 \pi T \delta g_{x^ix^i}^{(0)} + 2  k_0  \delta g_{vx}^{(0)} \bigg] = 0 \,.
\label{pert2}
\end{equation}
Note that even though we have only moved slightly away from the special point, Einstein's equation \eqref{pert2} now imposes a non-trivial constraint on the near-horizon expansion that must be satisfied by the ingoing mode (in addition to those coming from the smooth limits of $X=0$). As such for a fixed $\delta \omega/\delta k$ there is no longer an extra solution. However, the singular nature of \eqref{location} is reflected in the fact that the constraint \eqref{pert2} depends on the slope $\delta \omega/\delta k$ with which we move away from \eqref{location}. This means that near \eqref{location} then, even after fixing the usual independent parameters in the near-horizon expansion, we have a family of different ingoing modes parameterised by $\delta \omega/\delta k$. The slope $\delta \omega/\delta k$ now acts as an extra parameter in the near-horizon solution and can be tuned to match the ingoing solution onto different asymptotic solutions at the boundary.

In particular, by choosing $\delta\omega/\delta k$ appropriately, we can ensure that  we have an ingoing mode that matches continuously onto the normalisable solution at the boundary. This can be achieved by choosing $\delta\omega/\delta k$ to satisfy %
\begin{equation}
\label{slope1}
\frac{\delta \omega}{\delta k} = \frac{ 2  k_0 \delta g_{vv}^{(0)}}{  \frac{d}{2}h'(r_0) \delta g_{vv}^{(0)} -  2 \pi T  \delta g_{x^ix^i}^{(0)}- 2   k_0  \delta g_{vx}^{(0)}},
\end{equation}
where $\delta g_{\mu \nu}^{(0)}$ corresponds to the near-horizon expansion of the normalisable solution. 
 
We therefore deduce that if we move away from the special point \eqref{location} along the slope \eqref{slope1}, then we will see a line of poles in the energy density correlation function that passes through \eqref{location} (as well as in the correlation functions of the operators dual to the fields that couple to $\delta g_{vv}$). Furthermore, given knowledge of the normalisable solution of the perturbation equations  then \eqref{slope1} gives a prediction for the slope of this line of poles in terms of the metric components of this mode. Note that we could also choose a different slope so that the ingoing mode matches onto a solution with different asymptotics at the boundary. In particular one can choose the slope to instead match to a solution with the expectation value $\langle T_{00}(\omega,k) \rangle_{\beta_0} = 0$, from which we also deduce the existence of a line of zeroes  in the dual Green's functions passing through \eqref{location}, with a different slope that is again related to the metric components of this solution through the same equation \eqref{slope1}.
 
Whilst the above discussion gives a simple gravitational argument for the phenomenon of pole-skipping for a broad class of black brane solutions, it is somewhat abstract. {In particular, although the existence of an extra parameter in the ingoing solution should generically result in an extra parameter in the asymptotic behaviour of $\delta g_{vv}$, the precise dependence of the asymptotic $\delta g_{vv}$ on the extra parameter (i.e.~the slope $\delta\omega/\delta k$) depends on the details of the specific gravitational theory.} To show how the argument works more directly, in the next section, we examine in detail a holographic model in which we can analytically study the energy density two-point Green's function. We will see explicitly all the features mentioned above, and confirm the prediction~\eqref{slope1}. Note that the above discussion applies in the vicinity of Eq. \eqref{location}, thus only showing that a line of poles passes through~\eqref{location}. It is not clear that the line of poles actually goes over to those that correspond to hydrodynamic excitations at small $\om$ and $k$. This will be checked by using explicit examples in the next section.

\section{Energy density Green's function near special point }
\label{sec:greensfunction}

In this section, we study the connection between chaos and the energy density correlator in a specific holographic model \cite{Andrade:2013gsa,Davison:2014lua}, which has a free parameter $m$ controlling the strength of momentum dissipation. We choose to study this example because it provides a simple and soluble model, which nevertheless exhibits a wealth of transport behaviour as one varies $m$, thus allowing us to demonstrate explicitly the insensitivity of the pole-skipping phenomenon to long distance transport properties of a system. Remarkably, in this model, we will be able to analytically compute the retarded energy density correlator near the special point \eqref{location} in Fourier space. This will allow us to demonstrate explicitly how the general considerations of Section~\ref{sec:generalargument} are borne out in a specific example. In particular, we will be able to prove that both the line of poles and the line of zeroes of $G^{R}_{T^{00} T^{00}}(\omega,k)$ always pass through \eqref{location} in these models. Moreover, we will be able to derive an analytic expression for the slope of the line of poles (of the dispersion relation $\omega(k)$) as it passes through the point \eqref{location}, which we find to be in perfect agreement with a numerical calculation of the same quantity. In Appendix~\ref{app:ads5}, we show that the same analytic discussion can also be applied to the Schwarzschild black hole of AdS$_5$, dual to a four-dimensional CFT at a finite temperature and studied initially in \cite{Grozdanov:2017ajz}.

Here, we focus on the $2+1$ dimensional boundary theories dual to $3+1$ dimensional gravity coupled to massless scalar (`axion') fields with action
\begin{equation}
S=\int d^4x\sqrt{-g}\left(\mathcal{R}+6-\frac{1}{2}\sum_{i=1}^{2}\partial_\mu\varphi_i\partial^\mu\varphi_i\right).
\label{action}
\end{equation}
In ingoing EF coordinates, this action has the black brane solution \cite{Andrade:2013gsa}
\begin{equation}
\begin{aligned}
\label{eq:BGsolution}
ds^2&=-r^2f(r) dv^2 + 2 dv dr +  r^2dx_idx^i\, , \qquad\qquad  f(r)=1-\frac{m^2}{2r^2}-\left(1-\frac{m^2}{2r_0^2}\right)\frac{r_0^3}{r^3}\, ,\\
\varphi_i &=mx^i \,.\\
\end{aligned}
\end{equation}
When $m=0$, \eqref{eq:BGsolution} is simply the Schwarzschild-AdS$_4$ solution and our results in this limit are identical to those that would be found in the absence of matter. Increasing $m$ causes the scalar fields to backreact on the metric, and in the extreme limit $m/T\rightarrow\infty$ the metric near the horizon becomes AdS$_2\times R^2$. 

The parameter $m$ sets the strength of the source of a scalar operator $\phi_{i}^{(0)}=mx_i$ that explicitly breaks the translational symmetry of the dual field theory. This symmetry breaking radically alters how energy is transported over long distances, resulting in a much richer variety of energy dynamics than in the translationally invariant case \cite{Davison:2014lua}. 

As we have explained, our goal is to study the energy density correlator of these theories near the point \eqref{location} in Fourier space. Explicitly, the location of this point is (see equations \eqref{location} and \eqref{eq:location0})
\begin{align}
\label{locationexplicit}
\omega = i  \lambda\,, && \lambda = 2 \pi T= \frac{r_0^2 f'(r_0)}{2}  \,,&&  k = i k_0 \,, && k_0^2 =  r_0^3 f'(r_0) \,,
\end{align}
and depends explicitly on the dimensionless parameter $m/T$.

We emphasise that all of our results hold even in the translationally invariant $m=0$ case. We allow for $m\ne0$ simply to illustrate the generality of the pole-skipping phenomenon, and its insensitivity to the long distance transport properties of the theory. Analogous results to those we will present can also be found in higher dimensions, and in Appendix~\ref{app:ads5} we outline how to obtain these for the Schwarzschild-AdS$_5$ solution. This solution was studied numerically in \cite{Grozdanov:2017ajz}, and our analysis provides an explanation of the results found there.

\subsection{Master field perturbation equations} 

In order to calculate the energy density correlation function we need to study the linearised gravitational perturbation equations about the spacetime  \eqref{eq:BGsolution}. In a general theory, studying these linearised perturbation equations is rather complicated, since at non-zero frequency $\omega$ and momentum $k$ there are many coupled fields. For instance, if one aligns the momentum in the $x$ direction then for this axion model, the correlator $G^R_{T^{00} T^{00}}(\omega,k) $ can be extracted from solving the equations of motion for the `longitudinal' perturbations $\{\delta g_{vv}, \delta g_{xx}, \delta g_{yy},\delta g_{rr},\delta g_{vx},\delta g_{xr},\delta g_{vr},\delta \varphi_1\}$ after Fourier transforming $\delta g_{\mu \nu} = \delta g_{\mu \nu}(r) e^{- i \omega v + i k x}$.

Fortunately, in the simple model \eqref{action} the dynamical equations for the perturbations can be decoupled by working with suitable gauge-invariant `master field' variables \cite{Davison:2014lua}. In particular, the energy density correlation function is simply related to the dynamics of the following variable
\begin{eqnarray}
 \label{eq:psidef} 
\psi \equiv   &\,\,r^4& f \bigg\{ \frac{d}{dr} \bigg[ \frac{\delta g_{xx}+ \delta g_{yy}}{r^2} \bigg]  - \frac{i \omega}{r^4 f}( \delta g_{xx} + \delta g_{yy} ) -\frac{2ik}{r^2}\bigg(\delta g_{xr} + \frac{\delta g_{vx}}{r^2 f} \bigg) \nonumber \\
&-& 2 r f \bigg(\delta g_{rr}+ \frac{2}{r^2 f} \delta g_{vr} + \frac{1}{r^4 f^2} \delta g_{vv} \bigg) - \frac{k^2+r^3f'}{r^5 f} \delta g_{yy} \bigg\} \nonumber \\
&-&\frac{\left(k^2+r^3f'\right)}{\left(k^2+m^2\right)}\frac{m r}{2}\bigg(\frac{m}{r^2}(\delta g_{xx}-\delta g_{yy})-2ik \delta\varphi_1\bigg) ,
\end{eqnarray}
which obeys the equation of motion
\begin{equation}
\label{eq:psieom}
\frac{d}{dr}\left[ \frac{r^2 f}{(k^2 + r^3 f')^2} \psi'\right] - \frac{2 i \omega}{(k^2 + r^3 f')^2} \psi' - \frac{ \Omega(\omega,k) }{r^2\left(k^2+r^3f'\right)^3} \psi=0 \,,
\end{equation}
where primes denote derivatives with respect to $r$ and we have defined
\begin{equation}
\Omega(\omega,k) =  ( 2 r_0^2 - m^2) 3 i r_0 \omega + k^2 (k^2 + m^2)  .
\end{equation}
The retarded two-point Green's function of the energy density $T_{00}$ is determined by solving \eqref{eq:psieom} subject to ingoing boundary conditions at the black hole horizon. From the near-boundary expansion of the solution $\psi=\psi^{(0)}+\psi^{(1)}r^{-1}+\ldots$ one can then read off the retarded Green's function using the usual holographic dictionary as
\begin{equation}
\label{eq:retardedgreenspsi}
G^R_{T^{00} T^{00} }(\omega,k)= k^2\left(k^2+m^2\right) \frac{\psi^{(0)}(\omega,k)}{\psi^{(1)}(\omega,k) + i \omega \psi^{(0)}(\omega,k)} \, , 
\end{equation}
up to contact terms.

As motivated earlier, we are particularly interested in demonstrating that there is always a line of poles $\omega(k)$ 
in \eqref{eq:retardedgreenspsi} that passes through \eqref{location}. Since we are working in terms of the master field \eqref{eq:psidef}, which is related to metric components in a complicated way, the usual notion of normalisable and non-normalisable modes is a bit subtle in terms of $\psi$. However, we can see that \eqref{eq:retardedgreenspsi} will have a pole when there is an ingoing solution with  $\psi^{(0)}(\omega, k) \neq 0$  and $\psi^{(1)}(\omega,k) + i \omega \psi^{(0)}(\omega,k) = 0$. It can be checked that a solution with these properties corresponds to a normalisable metric perturbation, and we will refer to it as a normalisable mode $\psi_{n}$. 

Likewise, there will be a line of zeroes passing through \eqref{location}, if near \eqref{location} we can show there is an ingoing mode with no normalisable component: $\psi^{(0)}(\omega, k) = 0$ and $\psi^{(1)}(\omega,k)\neq 0$. For \eqref{eq:retardedgreenspsi} to have both a line of poles and zeroes arbitrarily close to \eqref{location} we therefore require there to be two different ingoing solutions as we approach \eqref{location}. In Section~\ref{sec:generalargument} we argued that indeed there is not a unique ingoing solution near \eqref{location}. We will now show how this same phenomenon can be seen directly from solving the equation of motion \eqref{eq:psieom} for the gauge invariant field $\psi$. 

\subsection{Solutions at special point}
\label{sec:specialpoint}

The existence of multiple different solutions near the special location \eqref{location} in Fourier space can be understood clearly by examining the near-horizon behaviour of \eqref{eq:psieom} at \eqref{location}. First, let us consider the near-horizon behaviour of \eqref{eq:psieom} at generic $k$. Then taking the near-horizon limit of \eqref{eq:psieom} one finds the following power law solutions to \eqref{eq:psieom}:
\be \label{onn0}
\psi = a_1 \eta_1 + a _2 \eta_2 , \qquad \eta_{1,2} = (r-r_0)^{\al_{1,2}} \;\;{\rm as}\;\; (r \to r_0) , \;\;\;\;\;\;\;\;\; \al_{1} = \frac{i \omega}{2 \pi T}, \;\;\;\;  \al_{2} = 0.
\ee
At a generic value of $\omega, k$ there is therefore a single regular solution in ingoing coordinates, corresponding to the $(r - r_0)^0$ power law. The power law $(r - r_0)^{\frac{i \omega}{2 \pi T}} $ is the corresponding outgoing solution. 

However, we can see from \eqref{eq:psieom} that at the special value $k^2 = -k_0^2$ the near-horizon behaviour of \eqref{eq:psieom} changes since $\left(k^2 + r^3 f'\right)$ now vanishes at the horizon. Taking the near-horizon limit of \eqref{location} one now instead finds power law solutions of the form
\be \label{onn01}
\psi = a_1 \eta_1 + a _2 \eta_2 , \qquad \eta_{1,2} = (r-r_0)^{\al_{1,2}} \;\;{\rm as} \;\; (r \to r_0) , \;\;\;\;\;\;\;\;\; \al_{1} = 1 + \frac{i \omega}{2 \pi T},\;\;\;\;  \al_{2} = 1.
\ee
The indices are shifted by one at this special value of $k$. For a generic $\omega$ it is clear that there is still one regular solution in ingoing coordinates corresponding to the $(r - r_0)$ power law. However, one can now see that something special occurs when we also set $i \omega = - 2 \pi T$ in \eqref{onn01}. In this case we have power law solutions
\be \label{onn1}
\psi = a_1 \eta_1 + a _2 \eta_2 , \qquad \eta_{1,2} = (r-r_0)^{\al_{1,2}} \;\;{\rm as} \;\; (r \to r_0) , \;\;\;\;\;\;\;\;\; \al_{1} = 0, \;\;\;\;  \al_{2} = 1.
\ee
As such, precisely at the location \eqref{location} it appears that both near-horizon solutions for $\psi$ are regular in ingoing coordinates. 

The existence of two regular solutions for $\psi$ at this point can be seen more directly from the fact that in this simple example we can exactly solve \eqref{eq:psieom} at the special point \eqref{location}. At this location $\Omega(\omega,k) = 0$ and hence \eqref{eq:psieom} simplifies considerably to
\begin{align}
\label{eq:psieom22}
\frac{d}{dr}\left[ \frac{r^2 f}{(-k_0^2 + r^3 f')^2} \psi'\right] + \frac{ r_0^2 f'(r_0)}{(-k_0^2 + r^3 f')^2} \psi' =0.
\end{align}
This is a first-order equation for $\psi'$ and so it is straightforward to find the general solution,
\begin{equation}
\label{eq:uvsol}
\psi(r)= c_1 + c_2 \int_{r_0}^{r} \mathrm{d}r \frac{\mathrm{exp}\bigg(\frac{m^2 - 3 r_0^2}{r_0 \sqrt{3r_0^2 - 2 m^2}} \mathrm{tan^{-1}}\bigg({\frac{2 r + r_0}{\sqrt{ 3 r_0^2 - 2 m^2}} \bigg) \bigg)}}{r \sqrt{2(r^2 + r r_0 + r_0^2)- m^2 }}\,.
\end{equation}
Note that the integral in \eqref{eq:uvsol} is regular as $r \to r_0$. We therefore manifestly have regular solutions near the horizon for any values of $c_1, c_2$, and these solutions can be expanded as 
\begin{equation}
\label{eq:uvsolexp}
\psi(r)= c_1 + c_2 F(r_0)(r - r_0) + \dots,  \;\;\;\;\;\;\;\;\;  F(r_0) =  \frac{\mathrm{exp}\bigg(\frac{m^2 - 3 r_0^2}{r_0 \sqrt{3r_0^2 - 2 m^2}} \mathrm{tan^{-1}}\bigg({\frac{3 r_0}{\sqrt{ 3 r_0^2 - 2 m^2}} \bigg) \bigg)}}{r_0 \sqrt{6 r_0^2 - m^2 }}.
\end{equation}

So precisely at \eqref{location} there is indeed an extra regular solution for $\psi$. That is, there are two linearly independent solutions to \eqref{eq:psieom22} that are both regular at the horizon and as such we can always choose $c_1$ and $c_2$ independently in \eqref{eq:uvsol}. Expanding the solution \eqref{eq:uvsol} near the boundary, it is clear that $\psi^{(1)}/\psi^{(0)}$, the ratio of coefficients that determines the Green's function \eqref{eq:retardedgreenspsi},  depends on $c_2/c_1$ and thus is not uniquely fixed by imposing ingoing boundary conditions at the horizon. By appropriately choosing $c_2/c_1$ one can find both normalisable and non-normalisable solutions that are regular at the horizon. In order to resolve this non-uniqueness, it is necessary to move slightly away from the location \eqref{location}, at which point the near-horizon behaviour reduces to \eqref{onn0}. We will do this shortly. 

We note that the considerable simplification $\Omega(\omega,k)=0$ that occurs at the location \eqref{location} does not generalise to higher dimensions. It is this simplification that allowed us to write down the exact expression \eqref{eq:uvsol} for the general solution for $\psi$. However, we emphasize that a simplification like this is not necessary to realise the key property that there are two independent solutions for $\psi$ that are regular at the horizon. In Appendix \ref{app:ads5}, we discuss the case of Schwarzschild-AdS$_5$ and illustrate this property by constructing regular series solutions for $\psi$ near the horizon.  

\subsection{Solutions near special point} 
\label{sec:analyticgreensfunction}

In order for the choice of ingoing boundary conditions on $\psi$ to be non-trivial, it is necessary to move slightly away from the location \eqref{location} in Fourier space. To do this we will consider perturbing a small distance $\epsilon$ from \eqref{location} by taking 
\begin{align}
\label{eq:scaling1}
k^2 = - k_0^2 + \ep ,&&  \omega = i \lambda  -  2 i  \lambda \epsilon q,
\end{align}
where we have introduced $q$ as a convenient parameterisation of the direction $\delta \omega/\delta k$ in which we move away from \eqref{location}
\begin{equation}
\label{gty}
\frac{\delta \omega}{\delta k} = \frac{4 \lambda^2}{{v_B}} q + {\cal O}(\epsilon).
\end{equation}
Note that near the horizon, the function appearing in the equation \eqref{eq:psieom} for $\psi$ takes the form
\begin{align}
\label{eq:regime}
k^2 + r^3 f'(r) = \ep + b (r-r_0) + O((r-r_0)^2), && b = 3 r_0^2 f'(r_0) + r_0^3 f''(r_0)  ,
\end{align}
and so the effects of $\epsilon\ne0$ on the equation become significant in the regime $(r - r_0) \sim \epsilon/b $. In particular for $(r - r_0) \ll \epsilon/b$ then the near-horizon behaviour reduces to \eqref{onn0} and the ingoing boundary condition for $\psi$ can be imposed near the horizon in the usual way.\footnote{Note that an exception to our analysis is provided by the special case $m^2 = 2 r_0^2$. In this axion model one finds $b = (3 m^2 - 6 r_0^2)/(2 r_0)$ and so the parameter $b$ vanishes for this choice of $m$. The point $m^2 = 2 r_0^2$ corresponds to a special point in this model in which there is an enhanced symmetry, and is discussed separately in Sec.~\ref{sec:sl2r} and Appendix~\ref{app:SL2}.}
\begin{figure}
\begin{center}
\resizebox{100mm}{!}{\includegraphics{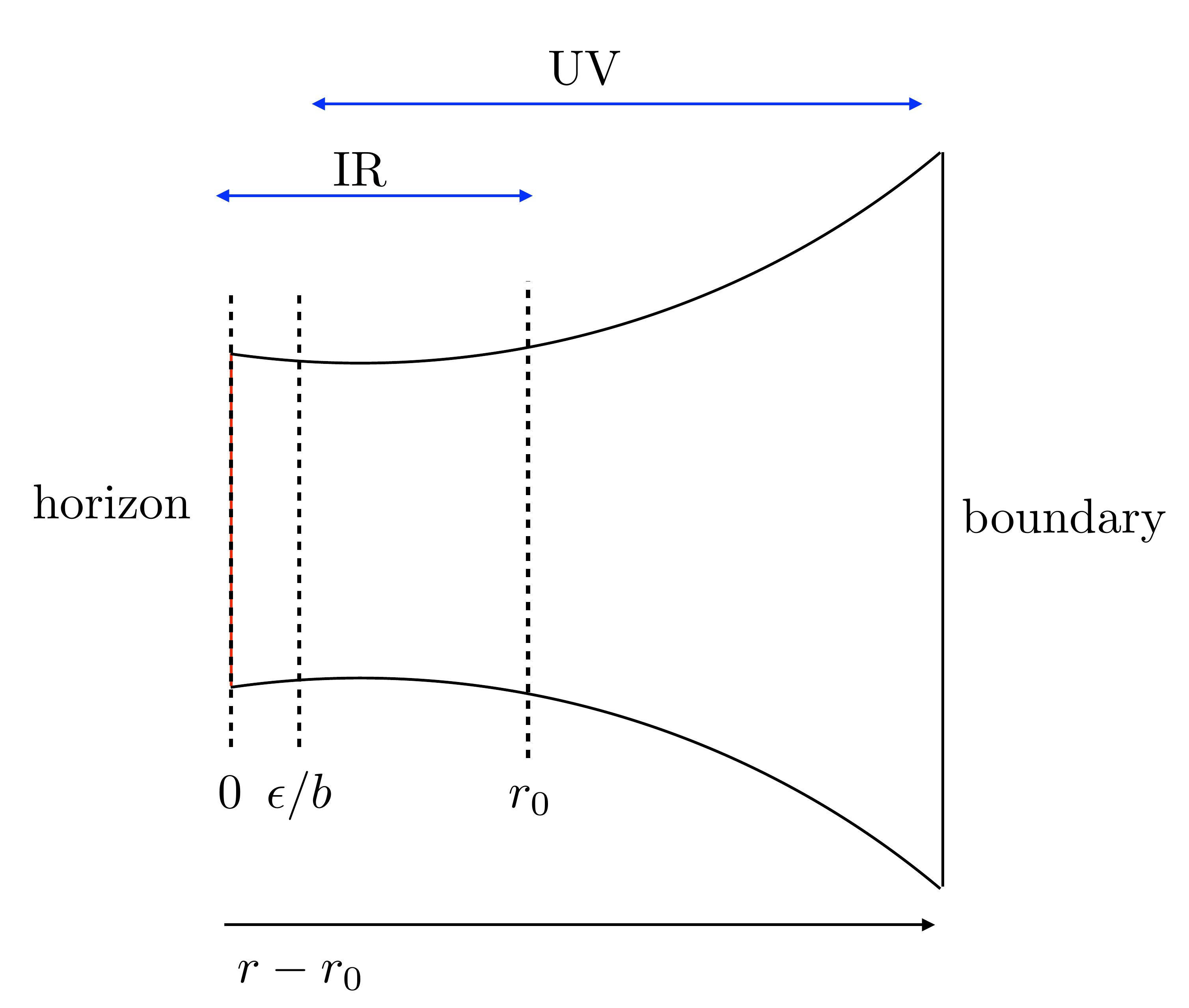}}
\caption{Moving slightly away from \eqref{location} changes the near-horizon behaviour of $\psi$. To calculate the Green's function we solve separately for the solution in the UV region $ (r - r_0) \gg  \epsilon/b$ and an IR region $(r - r_0) \sim \epsilon/b$. The solutions can then be matched by comparing them in their overlapping regime of validity.}
\label{fig:matching}
\end{center}
\end{figure}

Our goal now is to construct the form of this ingoing solution as we take $\ep \to 0$ and approach the point \eqref{location}. We can do this by dividing the radial direction into two regions and performing a matching calculation (see Figure~\ref{fig:matching}). In particular for $(r - r_0) \gg \ep/b$ we are free to safely ignore $\ep$ in \eqref{eq:psieom}. In this `UV regime' the solutions are then given by solving \eqref{eq:psieom} at $\omega = i \lambda$ and $k = i k_0$ and simply take the form \eqref{eq:uvsol}. 

However, close to the horizon $(r - r_0) \sim \ep/b$ the approximation of ignoring $\ep$ breaks down. We therefore need to solve separately for the solution in this `IR regime'. To do this we can consider the scaling  
\be 
\label{eq:scaling2}
(r-r_0) = \frac{\ep y}{b}, 
\ee
and then construct the ingoing solution perturbatively in $\epsilon$ 
\begin{equation}
\label{eq:scaling3}
\psi_{r}(y) = \psi_{0}(y) + \epsilon \psi_{1}(y) + \dots.
\end{equation}
By expanding this solution in the regime $y \gg1 $ we will then obtain a solution of the form \eqref{onn1} that can be matched to the UV solution by comparing to \eqref{eq:uvsolexp}. 

Specifically we insert \eqref{eq:scaling1}, \eqref{eq:scaling2} and \eqref{eq:scaling3} into the equation of motion for $\psi$ and expand in small $\epsilon$. At leading order \eqref{eq:psieom} becomes (primes now denote derivatives with respect to $y$)
\be \label{2}
\psi_0'' + \frac{2}{y(1 + y)} \psi_0'= 0 \ .
\ee
The regular solution to \eqref{2} simply corresponds to a constant
\begin{equation}
\psi_0(y) = c_1,
\end{equation}
which we can set to unity $c_1 = 1$ to normalise our solution. At this order in our expansion the regular solution therefore only has one of the power laws in \eqref{onn1}. To match to the UV we also need to determine the coefficient of the $y^1$ term. This requires us to go to next order in our expansion \eqref{eq:scaling3} and solve for $\psi_1(y)$. At ${\cal O}(\cal \epsilon)$ we find that the equation of motion takes the form
\be
\psi_1'' + \frac{2}{y(1 + y)} \psi_1' = f_1(y, m^2, q, r_0),
\ee
where the forcing term $f_1(y,m^2,q, r_0)$ is determined by expanding the equation of motion and then inserting the zeroth-order solution $\psi_0 = 1$. We can then again solve this equation subject to regularity at the horizon to find $\psi_1(y)$. Expanding this solution at large $y$ we find that the leading behaviour of $\psi_1$ is
\begin{equation}
\psi_1(y) = b_+(q) y,
\end{equation}
where $b_+(q)$ is explicitly given by the formula
\begin{align}
\label{bplus0}
b_+(q) = b_0 -  q, && b_0 = \frac{4(3 r_0^2 -  m^2)}{3(6 r_0^2 - m^2)(2 r_0^2 - m^2)}. 
\end{align}
After transforming back to the radial coordinate $r$ we then find that the ingoing solution takes the form 
\begin{equation}
\label{irsolution}
\psi_r(r) =  1+  b_+(q) b (r - r_0) + \dots,  
\end{equation}
in the IR region, where the $\dots$ indicate terms that vanish as $\epsilon \to 0$.\footnote{Note that even though the $y^{1}$ power law arose from $\psi_1$ it appears at leading order in \eqref{irsolution} as it carried a different $y$ dependence than the zeroth-order solution.}

The key point is that as we take $\ep \to 0$ the ingoing mode in \eqref{irsolution} depends at ${\cal O}(\epsilon^0)$ on the direction $q$ at which we approach the special point \eqref{location}. In particular, by varying the direction $q$ we can generate an arbitrary linear combination of the two solutions in \eqref{onn1}. By choosing an appropriate $q$, we can therefore always match the ingoing solution onto any UV solution~\eqref{eq:uvsolexp}. Explicitly performing this matching we find that the ingoing solution \eqref{irsolution} matches to the UV solution
\begin{align}
\label{uvsol}
\psi_{\mathrm{UV}}(r)  =  1 + \frac{b_+(q) b }{F(r_0)} \int_{r_0}^{r} \mathrm{d}r \frac{\mathrm{exp}\bigg(\frac{m^2 - 3 r_0^2}{r_0 \sqrt{3r_0^2 - 2 m^2}} \mathrm{tan^{-1}}{\frac{(2 r + r_0)}{\sqrt{ 3 r_0^2 - 2 m^2}} \bigg)}}{r \sqrt{2(r^2 + r r_0 + r_0^2)- m^2) }}.
\end{align}

In particular, since we can match to any UV solution by a suitable choice of $q$, it is clear that there is always a line of poles passing through the location \eqref{location}: by choosing the direction $q$ at which we move away from the special point appropriately, we will find an ingoing solution which is normalisable in the UV ($\psi_n^{(0)} = 1, \psi_n^{(1)} - 2 \pi T \psi_n^{(0)} = 0$) and there will therefore be a line of poles passing through the point \eqref{location}. Moreover we can use our matching procedure to obtain a prediction for the slope of this line of poles passing through \eqref{location}. To do this we simply expand the normalisable solution $\psi_n$ as $r \to r_0$ as in \eqref{eq:uvsolexp} to find a solution of the form
\begin{equation}
\psi_{n} = a_{1,n} + a_{2,n} (r - r_0) + \dots, 
\end{equation}
with fixed coefficients $a_{1,n}$ and $a_{2,n}$ that we display in Appendix~\ref{app:matching}. To match $\psi_{n}$ with our infrared solution \eqref{irsolution} then we simply have to choose the slope $q = q_p$ such that 
\begin{equation}
\label{prediction}
b_+(q_p) b  = \frac{a_{2,n}}{a_{1,n}}. 
\end{equation}
With this choice of $q=q_p$, we then have that the ingoing mode in \eqref{irsolution} is normalisable in the UV. As such if one moves away from \eqref{location} along this direction then one will see a line of poles passing through \eqref{location} with this slope. Combining~\eqref{prediction},~\eqref{gty} and the explicit forms of $a_{1,n}$ and $a_{2,n}$ from Appendix~\ref{app:matching}, we then find an analytic formula for the slope of the line of poles passing through \eqref{location}:
\begin{equation}
\label{eq:explicitpredict}
\frac{1}{v_B} \frac{\delta \omega}{\delta k} = \frac{8(3 r_0^2 - m^2)}{3(2 r_0^2 - m^2) } + \frac{4 (6 r_0^2 - m^2 )}{3(m^2 - 2 r_0^2)} \frac{F(r_0) r_0}{{\tilde N}(m, r_0)  },
\end{equation} 
where $\tilde{N}$ can be written in terms of the integral of a known function and is defined through \eqref{N} and \eqref{eq:ntilde}. 

We have focused on describing the matching to $\psi_n$ since this establishes the existence of a pole passing through \eqref{location}. However, by moving away from the point \eqref{location} along a different direction $q$ we could find an ingoing solution that matches to \emph{any} UV solution. In particular we could also pick a (different) $q=q_z$ to use \eqref{prediction} to match to the coefficients $a_{1,nn}$ and $a_{2,nn}$ of a solution which has no normalisable component.  As such there will always also be a line of zeros of the Green's function passing through \eqref{location} at a slope $q=q_z$. For a general $q$ the UV solution is the linear combination of these modes given by \eqref{uvsol}.  In Appendix~\ref{app:matching} we extract the full Green's function from \eqref{uvsol} and verify it indeed has the form
\begin{equation}
\label{greensfunctionpert}
G^R_{T_{00} T_{00} }(\omega,k)  \propto \frac{\delta \omega - 4\lambda^2 q_z/v_B \delta k}{\delta \omega -4 \lambda^2 q_p/v_B \delta k} \,, 
\end{equation} 
in which one can explicitly see there are both a line of poles and a line of zeroes passing through \eqref{location}. Note that the precise slope of the line of zeroes $q_p$ could be altered by the inclusion of contact terms and in this sense, it is less robust than the trajectory of the line of poles $q_p$.

It is worth emphasising that the behaviour of the Green's function in \eqref{greensfunctionpert} is rather unusual. Firstly, since there is not a unique ingoing solution at \eqref{location}, the Green's function in \eqref{greensfunctionpert} is not uniquely defined at this point. In particular, depending on how we approach \eqref{location} we may see a line of zeroes along $q = q_z$ or a line of poles along $q = q_p$. The fact that both a line of zeroes and a line of poles cross at \eqref{location} has been emphasised in \cite{Grozdanov:2017ajz,Blake:2017ris} and is referred to as pole-skipping.

\subsection{Expansion of $\psi$ at special point in terms of the metric}
\label{sec:metriccomparison}
So far we have seen two different approaches to describing the special nature of \eqref{location}. In Section~\ref{sec:generalargument} we used the Einstein equation in ingoing EF coordinates to argue that there was not a unique ingoing solution near \eqref{location}. In Section~\ref{sec:greensfunction} we explicitly saw this was the case by solving the equation of motion for the gauge invariant variable $\psi$ near \eqref{location}. Here we will use the relationship \eqref{eq:psidef} between $\psi$ and the metric to show these two descriptions are in precise agreement with one another.

To compare these discussions we can simply insert the near-horizon expansion 
\begin{eqnarray}
\delta g_{\mu \nu}(r) &=& \delta g_{\mu \nu}^{(0)} + \delta g_{\mu \nu}^{(1)} (r - r_0) + \dots, \nonumber \\ 
\delta \varphi_1(r) &=& \delta \varphi_1^{(0)} + \delta \varphi_1^{(1)} (r - r_0) + \dots, 
\end{eqnarray}
into the definition \eqref{eq:psidef} of the gauge invariant field $\psi$. This then allows us to construct the near-horizon behaviour of $\psi$ in terms of the expansion of a metric \eqref{eq:irexp} which is regular in ingoing coordinates. Upon doing so we find that at the special point \eqref{location} this gives a solution of the form
\begin{equation}
\label{expapp}
\psi = a_{1}  + a_{2} (r - r_0)  + \dots, 
\end{equation}
where $a_{1}$ and $a_{2}$ are related to the coefficients of the metric and $\delta \varphi_1$ in the near-horizon expansion \eqref{eq:irexp}. The coefficient $a_{1}$ has a simple form and can immediately be read off from the near-horizon expansion of the metric as 
\begin{equation}
\label{aminus}
a_{1} =  -\bigg[ {  2 r_0 \delta g_{vv}^{(0)} -  2 \pi T  \delta g_{xx}^{(0)} - 2 \pi T  \delta g_{yy}^{(0)} - 2   k_0  \delta g_{vx}^{(0)}} \bigg]. 
\end{equation} 
Determining the coefficient $a_{2}$ is more complicated as after inserting our expansion \eqref{eq:irexp} into \eqref{eq:psidef} one finds it depends on many different coefficients in the expansions of the near-horizon metric and scalar field. However since we are looking to determine $a_2$ for a solution to the equations of motion we can use Einstein's equations at \eqref{location} (the equations we schematically denoted $X=0$ in Section~\ref{sec:generalargument}) to simplify this. After doing so one finds the simple result
\begin{equation}
\label{aplus}
a_{2} = b_0 b a_{1} + \frac{b}{4\pi T} \delta g_{vv}^{(0)}, 
\end{equation}
with $b_0$ given by the formula in \eqref{bplus0} and where we have used the explicit expressions
\begin{align}
\label{eq:Texplicit}
4 \pi T = \frac{(6 r_0^2 - m^2) }{2 r_0} ,&& b = \frac{3(m^2 - 2 r_0^2)}{2 r_0}.
\end{align}

Now, as we emphasised in Section~\ref{sec:generalargument}, precisely when $\delta \omega =0$ and $\delta k = 0$, the $vv$ component of the Einstein equations is trivial and as a result the parameters $\delta g_{vv}^{(0)}, \delta g_{xx}^{(0)}, \delta g_{yy}^{(0)}, \delta g_{vx}^{(0)}$ can be chosen independently. As such when we are precisely at the point \eqref{location} the coefficients $a_1$ and $a_2$ in \eqref{expapp} are independent of each other. That is, at the special location \eqref{location} an arbitrary solution of the form \eqref{expapp} can be realised by a regular metric. The fact that there are two independent regular solutions for $\psi$ can be seen to be a direct consequence of the fact that \eqref{perteqn0} did not impose a constraint on the near-horizon metric, and hence we had an extra free parameter in the solution. 

However, slightly away from \eqref{location}, we know that these metric components are not independent, but rather are related through Einstein's equation \eqref{perteqn0}. From our matching argument we determined the solution \eqref{irsolution} for $\psi$ by imposing ingoing boundary conditions slightly away from the point \eqref{location}. This matching resulted in the value of $a_2/a_1$ being determined by the direction $q$ in which the special point \eqref{location} is approached: $b b_+(q) = a_2/a_1$. From using the expressions \eqref{aminus} and \eqref{aplus} we now see that this expression relates the coefficients of the near-horizon metric perturbations to the direction $q$ in which we move away from the special point
\begin{equation}
\frac{1}{4 \pi T} \frac{  \delta g_{vv}^{(0)}}{  2 r_0  \delta g_{vv}^{(0)} -  2 \pi T  \delta g_{xx}^{(0)} - 2 \pi T  \delta g_{yy}^{(0)} - 2   k_0  \delta g_{vx}^{(0)}}=q,
\end{equation}
where have used the expression for $T$ in \eqref{eq:Texplicit}. From using $\delta \omega/\delta k = 4 \lambda k_0 q$ one then can see this is precisely equivalent to the previous prediction \eqref{slope1} we derived in Section~\ref{sec:generalargument} by imposing the $vv$ component of Einstein's equations \eqref{pert2} close to the special point \eqref{location} in Fourier space.

\subsection{Comparison to numerics and hydrodynamic poles}
\label{sec:hydropoles}

In Section~\ref{sec:analyticgreensfunction}, we analytically demonstrated that in the holographic model \eqref{action} there is a pole in the energy density correlator passing through \eqref{location}, with a slope determined by \eqref{bplus0}, for any value of $m/T$. Furthermore, as a line of zeroes also passes through \eqref{location}, all of these models exhibit the phenomenon of pole-skipping at the point \eqref{location}. Here we will confirm these analytical results by comparing them to numerical computations of the Green's function poles (the quasinormal modes of the spacetime) in \eqref{eq:retardedgreenspsi}. These numerics also allow us to extract the full dispersion relation $\omega(k)$ of the pole that passes through \eqref{location}. From this we investigate the behaviour of this same pole at small $\omega,k$ and explore the connection with hydrodynamic modes seen in \cite{Blake:2017ris, Grozdanov:2017ajz}. 

In particular, let us recall that in the translationally invariant (pure gravity) case $m=0$, the hydrodynamic limit of the energy density Green's function \eqref{eq:retardedgreenspsi} is dominated by gapless sound modes. That is, at small $\omega$ and $k$ it exhibits hydrodynamic sound poles with dispersion relations \cite{Herzog:2002fn}
\begin{equation}
\label{ads4sound}
\omega(k)=\pm v_sk-i\frac{k^2}{8\pi T}+\ldots,
\end{equation}
where $v_s=1/\sqrt{2}$ is the speed of sound and $\ldots$ denote higher-order corrections in $k$ that in principle can be computed using higher-order hydrodynamics (see e.g. \cite{Grozdanov:2015kqa}). As for the analogous AdS$_5$ theory studied numerically in \cite{Grozdanov:2017ajz}, we observe that the full dispersion relation $\omega(k)$ of the mode which approaches \eqref{ads4sound} (with $+$ sign for Im$(k)>0$) at small $k$ satisfies the condition $\omega(ik_0)=i\lambda$ where $\lambda$ and $k_0=\lambda/v_B$ are independently defined by the formula \eqref{locationexplicit}.

\begin{figure}
\begin{center}
\resizebox{80mm}{!}{\includegraphics{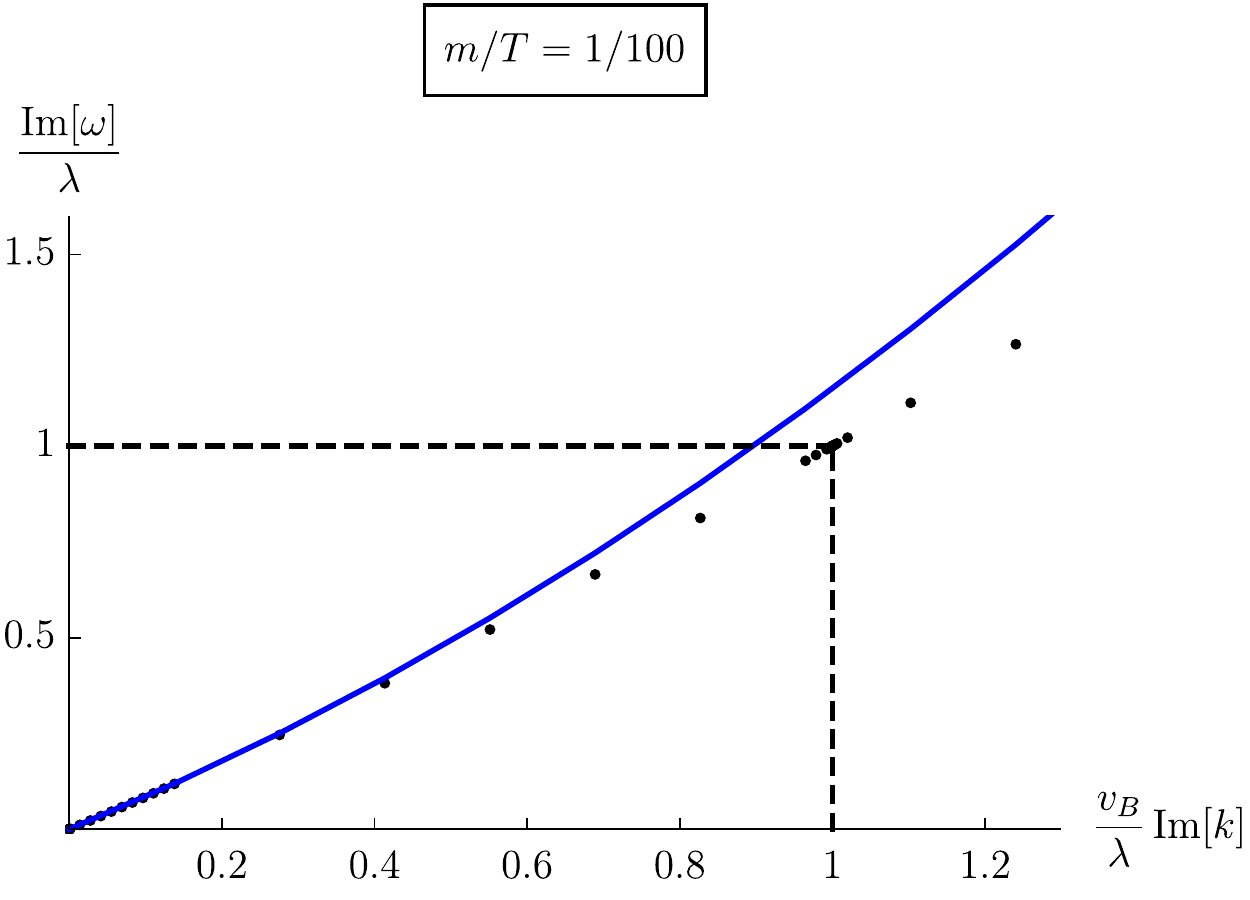}}
\hspace{8mm}
\resizebox{80mm}{!}{\includegraphics{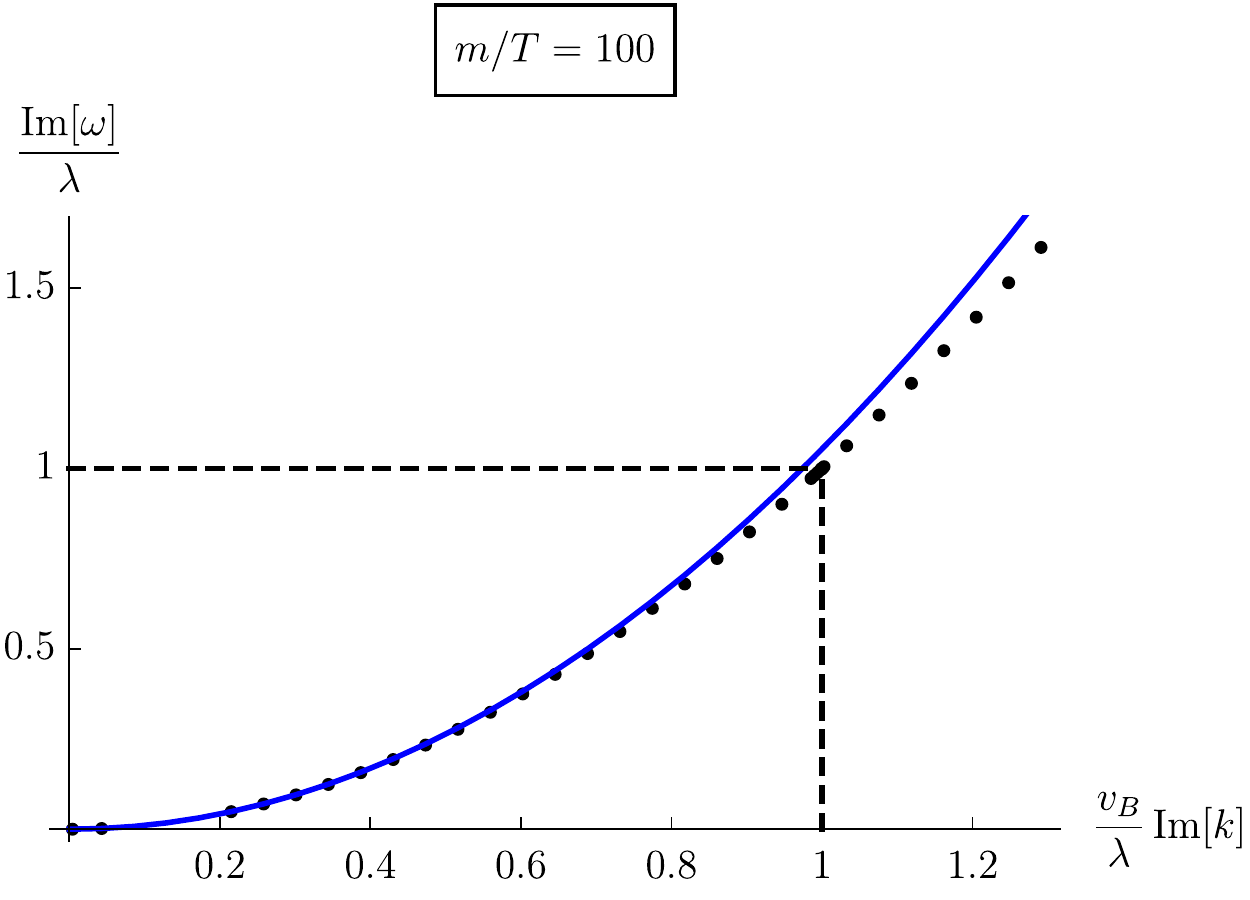}}
\caption{These plots show the dispersion relation $\omega(k)$ of the hydrodynamic pole in \eqref{eq:retardedgreenspsi} as a function  of imaginary $k$ for the choices $m/T =1/100$ and $m/T  =  100$. The blue lines are the hydrodynamic approximations \eqref{ads4sound} (left panel) and \eqref{ads4diffusion} (right panel) to the small $k$ hydrodynamic behaviour. The black dots correspond to the exact dispersion relation extracted from our numerics. Despite the qualitatively different small $k$ behaviour, in all cases we find this dispersion relation passes through the special point \eqref{location} such that $\omega(i k_0) = i \lambda$ for $k_0 = \lambda/v_B$.} 
\label{fig:plots}
\end{center}
\end{figure}

\begin{figure}
\begin{center}
\resizebox{80mm}{!}{\includegraphics{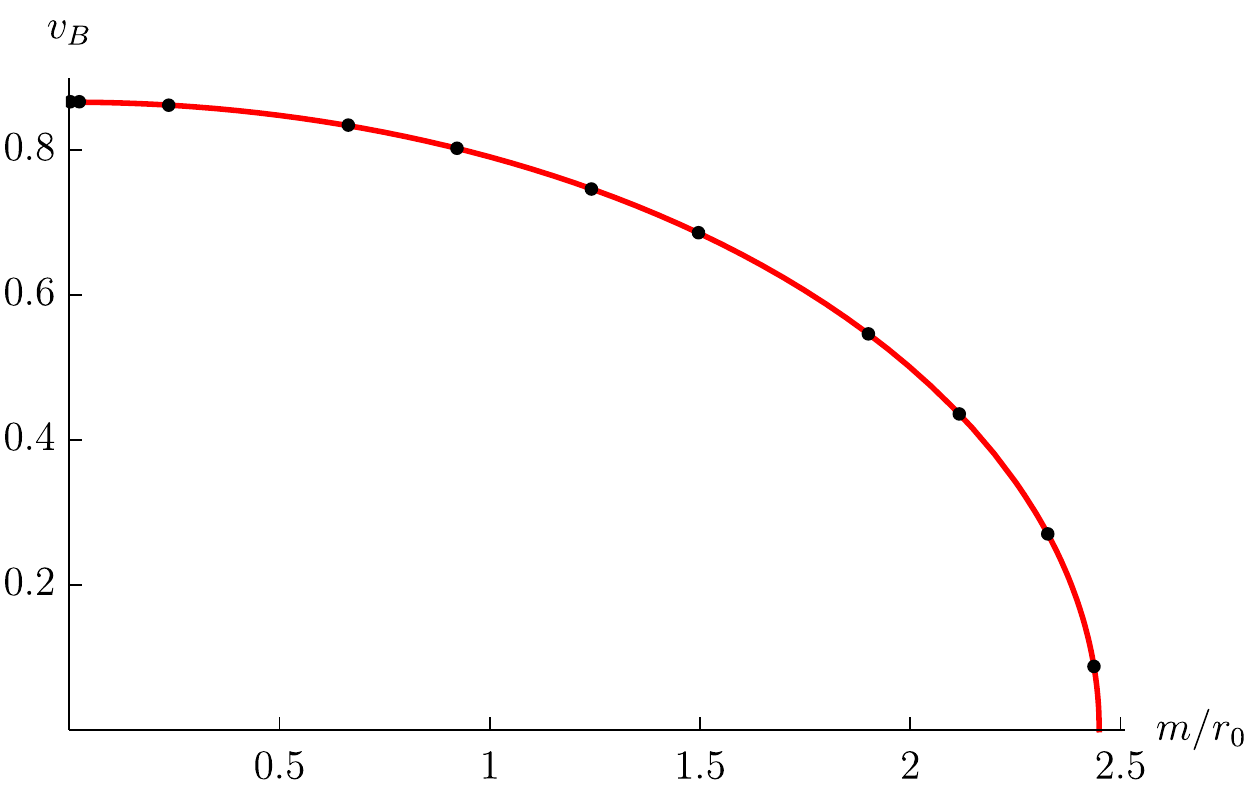}}
\hspace{8mm}
\resizebox{80mm}{!}{\includegraphics{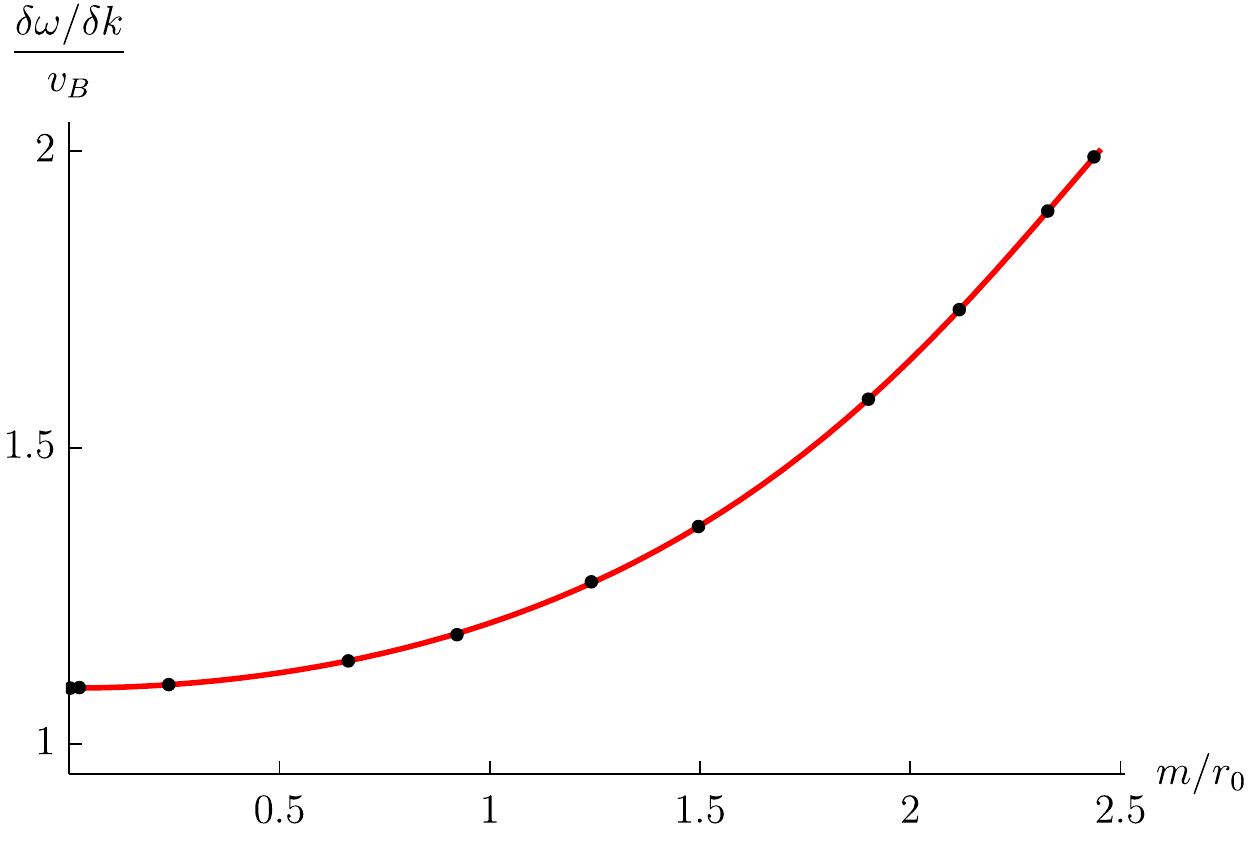}}
\caption{(a) The red line plots the butterfly velocity $v_B = \lambda/k_0$ as determined from the analytic formula \eqref{locationexplicit}. The black dots correspond to $\lambda/k_1$ where $k_1$ is extracted as the wavevector for which the numerical dispersion relation for the hydrodynamic pole satisfies $\omega(i k_1) = i \lambda$. If the pole passes through \eqref{location} then we should have $\lambda/k_1 =  v_B$ which indeed holds for all $m/T$. (b) The red line plots the analytic formula \eqref{eq:explicitpredict} for the slope of the line of poles in the energy density correlator as it passes through the point \eqref{location}. The dots correspond to the values of the slope extracted from the numerical calculations of the dispersion relation. Note that $m/T\rightarrow\infty$ corresponds to $m=\sqrt{6}r_0$, which is the upper limit shown on each plot.}
\label{fig:plot}
\end{center}
\end{figure}

Furthermore, we are able to demonstrate that this connection between hydrodynamic poles of the energy density Green's function and the location \eqref{location} is far more general, and does not rely on the hydrodynamic modes having a sound-like dispersion relation at small $k$. At any non-zero $m/T$, the relaxation of momentum qualitatively changes the nature of the hydrodynamic modes that propagate over long distances and one instead finds a diffusive pole in \eqref{eq:retardedgreenspsi} at small $k$ \cite{Davison:2014lua}
\begin{equation}
\label{ads4diffusion}
\omega(k)=-iD_Ek^2+\ldots,
\end{equation}
where $D_E$ is the energy diffusion constant. Numerically tracking the dispersion relation of this pole, we find that for any value of $m/T$ that $\omega(k)$ continues to pass through the location \eqref{location}. We emphasise that this is very non-trivial. There are a family of different dispersion relations $\omega(k)$, parameterised by $m/T$, whose qualitative features change dramatically as $m/T$ is varied (see Figure \ref{fig:plots}). However, in all cases we find that regardless of the details of the full dispersion relation it always satisfies the constraint $\omega(ik_0)=i\lambda$ as can be seen in the left hand panel of Figure \ref{fig:plot}.

In addition, from the numerical computations of the dispersion relation we can extract the slope $\delta\omega/\delta k$ of the pole as it passes through the location \eqref{location} and compare it to the analytic result \eqref{eq:explicitpredict} that we obtained from the matching calculation in Section \ref{sec:greensfunction}. The analytic formula \eqref{eq:explicitpredict} for $\frac{\delta\omega}{\delta k}/v_B$ interpolates between roughly $1.09$ for the translationally invariant case $m/T=0$ and exactly $2$ for $m/T\rightarrow\infty$. In the right hand panel of Figure \ref{fig:plot} we plot this formula (red line) and also the numerical values of the slope (black dots) extracted from the dispersion relation (e.g.~those shown in Figure \ref{fig:plots}). The numerical results agree perfectly with the analytic expression \eqref{eq:explicitpredict}, validating the details of the matching argument described in Section \ref{sec:greensfunction}.

As can be seen in Figure \ref{fig:plots}, as $m/T$ is increased, the leading-order hydrodynamic approximations of the dispersion relations $\omega(k)$ become a better and better approximation to the exact dispersion relation near the special point \eqref{location}. Using the fact that $D_E(m/T\rightarrow\infty)\rightarrow v_B^2/\lambda$ \cite{Blake:2016sud}, our result that $\frac{\delta\omega}{\delta k}/v_B\rightarrow2$ in this limit indicates that the hydrodynamic approximation to the dispersion relation \eqref{ads4diffusion} is exact in the vicinity of the point \eqref{location} when $m/T\rightarrow\infty$. This is consistent with the observation that hydrodynamic collective modes exist even over time scales much shorter than $T^{-1}$ in holographic theories with an AdS$_2$ factor in the near-horizon metric \cite{Davison:2013bxa}.

\subsection{The $SL(2,R) \times SL(2,R)$ invariant point} 
\label{sec:sl2r}

Our analytic discussion in Sections~\ref{sec:specialpoint} and~\ref{sec:analyticgreensfunction} describes the behaviour of the axion model near \eqref{location} at generic values of the parameters $m$ and $r_0$. However for the special choice of $m^2 = 2 r_0^2$ the above discussion in terms of the gauge invariant mode $\psi$ is somewhat subtle, as a result of the fact the parameter $b$ controlling our matching calculation vanishes for this choice of $m$. We perform a detailed analysis of this special case in Appendix~\ref{app:SL2}. For the purposes of our discussion in the main text, we simply note that at this value of $m/r_0$, the theory described by \eqref{action} dramatically simplifies and has an enhanced $SL(2,R) \times SL(2,R)$ symmetry \cite{Davison:2014lua}. This enhanced symmetry allows one to obtain an analytic expression for the Green's function $G^R_{T^{00} T^{00}}$ for any $\omega,k$. Here we will use this expression to demonstrate the pole-skipping phenomenon very explicitly. In particular, for this choice of $m^2 = 2 r_0^2$ the retarded Green's function takes the form\footnote{For other examples, which allow for exact solutions of holographic Green's functions at points of enhanced symmetry, see e.g. \cite{Compere:2012jk,Grozdanov:2016fkt,Bueno:2018xqc}.} (see Ref. \cite{Davison:2014lua})

\begin{equation}
G^R_{T^{00} T^{00}} (\omega , k ) = - \frac{k^2 \left(k^2 + 2 r_0^2 \right)}{2 r_0 } \frac{ \Gamma\left( \frac{1}{4} - \frac{i\omega}{2r_0} - \frac{1}{4} \sqrt{1 - 4 \frac{k^2}{r_0^2} }  \right) \Gamma\left( \frac{1}{4} - \frac{i\omega}{2r_0} + \frac{1}{4} \sqrt{1 - 4 \frac{k^2}{r_0^2} }  \right) }{\Gamma\left( \frac{3}{4} - \frac{i\omega}{2r_0} - \frac{1}{4} \sqrt{1 - 4 \frac{k^2}{r_0^2} }  \right) \Gamma\left( \frac{3}{4} - \frac{i\omega}{2r_0} + \frac{1}{4} \sqrt{1 - 4 \frac{k^2}{r_0^2} }  \right) }. 
\label{eq:exactcorrelator}
\end{equation}
From which we see that the Green's function has an infinite family of poles $\omega(k)$ satisfying
\begin{align}
\label{eq:specialpoles}
 \frac{1}{4} - \frac{i\omega}{2r_0} - \frac{1}{4} \sqrt{1 - 4 \frac{k^2}{r_0^2} } = - n , &&   \frac{1}{4} - \frac{i\omega}{2r_0} + \frac{1}{4} \sqrt{1 - 4 \frac{k^2}{r_0^2} }  = - p, 
\end{align}
where $n$, $p$ are non-negative integers. 
Expanding the pole with $n = 0$ gives the hydrodynamic mode
\begin{equation}
\omega = - i D_E k^2 + \dots,  
\end{equation}
with an energy diffusion constant $D_E = r_0^{-1}$. As in Section~\ref{sec:hydropoles} we can consider tracking this pole as we increase imaginary $k$. The full dispersion relation can be readily read from \eqref{eq:specialpoles}:
\begin{align}
\label{dispimaginary}
\omega =  i \, \frac{ r_0}{2} \left( \sqrt{1 - \frac{4 k^2}{r_0^2} } - 1 \right) ,
\end{align}
which we can see passes through $\omega = i \lambda = i r_0$ precisely at the momentum  
\begin{equation}
k^2 = - 2 r_0^2 = -k_0^2.
\label{eq:polewavevector}
\end{equation} 
Furthermore, from \eqref{dispimaginary}, we can also extract the slope of the pole as it passes through \eqref{location} as
\begin{align}
\frac{1}{v_B} \frac{\delta \omega}{\delta k} = \frac{4}{3},
\end{align}
which lies precisely on the curve derived from our matching argument~\eqref{eq:explicitpredict} and plotted in Figure~\ref{fig:plot}. Finally, we can see from \eqref{eq:exactcorrelator} that the Green's function also has a line of zeroes at $k^2 = - k_0^2$. The simple expression \eqref{eq:exactcorrelator} therefore exhibits all the expected features of pole-skipping. 

Whilst for the purposes of the main text we have simply focused on using \eqref{eq:exactcorrelator} to illustrate this phenomenon, the enhanced symmetry of the theory at $m^2 = 2r_0^2$ means we can perform a very detailed analysis of the origin of pole-skipping in this example. As we explain in Appendix~\ref{app:SL2}, there are subtleties involved in discussing pole-skipping in terms of the gauge invariant field $\psi$ at this point. Fortunately however this special case is sufficiently simple that we are able to also explicitly discuss this phenomenon in terms of the family of ingoing metric solutions discussed in Section~\ref{sec:generalargument}. In particular it is possible for this choice of $m$ to find exact analytic expressions for the metric solutions at \eqref{location} everywhere in the bulk, and hence study their UV asymptotics. As we discuss in Appendix~\ref{app:SL2} this allows us to confirm the existence of an extra parameter in the ingoing metric solution, and explicitly see that this extra parameter gives rise to the phenomenon of pole-skipping according to the discussion in Section~\ref{sec:generalargument}.

\section{Discussion}
\label{sec:discussion} 

In this paper we have shown that in general holographic models dual to Einstein gravity coupled to matter, remarkable signatures of many-body chaos exist in the energy density two-point correlation functions. A key element for the discussion is the observation that one of the Einstein's equations becomes trivial at the horizon at the special point~\eqref{location} determined by chaos parameters, which leads to a general argument for the phenomenon of pole-skipping \cite{Grozdanov:2017ajz,Blake:2017ris}.  We then illustrate how the general argument works 
in a specific holographic model in great detail. 

As emphasised in~\cite{Blake:2017ris}, the phenomenon of pole-skipping can be considered as a ``smoking-gun'' 
for the fact quantum many-body chaos is tied with energy conservation. The results of this paper give further strong support for this surprising, but likely profound connection (at least for maximally chaotic systems). In particular, from the perspective of the hydrodynamic dispersion relation~\eqref{poln}, the statement that $\om_h (k)$ must pass through~\eqref{location} is a highly nontrivial one. In the Einstein-Axion model we explicitly saw that this happened regardless of the detailed behavior of the full dispersion relation, which could vary dramatically as we changed the ratio $m/T$. The special point~\eqref{location} lies outside the range of the small $\om, k$ expansion and thus the full dispersion relation $\om_h (k)$ is needed to extrapolate to the point. This means that in the small $k$ expansion of the second equality of~\eqref{poln}  an infinite number of terms (which arise from higher-order hydrodynamics) must conspire for $\om_h (k)$ to pass through~\eqref{location}. In the proposal of~\cite{Blake:2017ris}, this conspiracy is ensured by an emergent shift symmetry in the hydrodynamic EFT for chaos. Thus, the results of this paper can also be considered as support for this ``hidden'' shift symmetry in hydrodynamics.

We end with a discussion of various questions for future research.

\ben

\item {\it Immediate generalizations }

The general argument we presented in Section~\ref{sec:generalargument} for pole-skipping implies that this phenomenon also occurs for charged black holes and/or for those with additional scalar fields, and it would be interesting to test this explicitly. Furthermore, whilst our analytic arguments show that there is a pole that always passes through \eqref{location}, it was only through the help of numerics that were able to see this pole to always be connected to the dispersion relation of the hydrodynamic mode. It would be interesting to understand if one can argue that this should always be the case from gravity, perhaps by tracking the holographic diffusion poles that were recently constructed using horizon data in \cite{Donos:2017ihe}.

\item {\it Shift symmetry in hydrodynamics from gravity} 

As commented above, the confirmation of pole-skipping in general holographic theories (which all have maximal chaos) can be 
considered as support for the shift symmetry in an all-order quantum hydrodynamics proposed in~\cite{Blake:2017ris}. It  would be extremely interesting to see whether the universal property of the linearised Einstein equation we uncovered at the special point~\eqref{location} can be used to identify this shift symmetry directly from a gravitational analysis. 

\item {\it Other implications of the extra ingoing mode}

It would be interesting to explore other physical implications of the extra ingoing mode we uncovered at the special point~\eqref{location}. For example, another important open question is to understand whether the existence of this pole-skipping phenomenon can be tied more directly to the existence of the Dray-`t Hooft shock wave solution that gives rise to \eqref{location}. In Section~\ref{sec:generalargument} we saw that the reason that pole-skipping occurred at \eqref{location} was a consequence of the fact that at $\omega = i \lambda$ the {Einstein's equation} \eqref{perteqn0} reduced to a decoupled equation for $g_{vv}^{(0)}$, that took the same form as the equation determining the shock wave profile \eqref{location}. However, as noted in \cite{Grozdanov:2017ajz}, there is not an immediate connection between such a perturbation and the delta function shock-wave in Kruskal-Szekeres coordinates. We hope to explore this question further in the future. 

\item{\it Higher-derivative gravity theories and stringy corrections} 

So far, all examples of pole-skipping are for maximally chaotic systems. It is clearly of crucial importance to understand
what happens to non-maximally chaotic systems. For holographic systems, stringy corrections decrease the Lyapunov exponent from being maximal~\cite{Shenker:2014cwa}, so it is very interesting to see how stringy corrections affect the pole-skipping. Perhaps a more immediate question is to see whether the phenomenon persists or gets modified when higher-derivative corrections are included. For example, Gauss-Bonnet gravity (see \cite{Brigante:2007nu,Grozdanov:2016vgg,Grozdanov:2016fkt}) may be a good testing ground for this purpose. 

More generally, it is also important to study this phenomenon in weakly coupled systems. A recent discussion~\cite{Grozdanov:2018atb} of chaos using kinetic theory gives some encouraging indications on the connection between chaos and energy dynamics even at weak coupling. 

\item{\it Theories with weak energy dissipation} 

Finally, it would be interesting to explore what happens if a system no longer has exact energy conservation. 
In the holographic context, weak energy dissipation can be introduced by working with massive gravity with an appropriate choice of graviton mass~\cite{Vegh:2013sk}. This, or axion models with time-dependent fields, could provide a nice laboratory to study how the Lyapunov exponent and pole-skipping are affected by energy dissipation.

\een

\neww{}

\vspace{0.2in}   \centerline{\bf{Acknowledgements}} \vspace{0.2in}

We are grateful to Koenraad Schalm and Vincenzo Scopelliti for valuable discussions. This work is supported by the Office of High Energy Physics of U. S. Department of Energy under grant Contract Number  DE-SC0012567. R.~D.~is supported by the Gordon and Betty Moore Foundation EPiQS Initiative through Grant GBMF\#4306. S. G. was supported by the U. S. Department of Energy under grant Contract Number DE-SC0011090. H. L. would also like to thank  Galileo Galilei Institute for Theoretical Physics for the hospitality during the workshop ``Entanglement in Quantum Systems'' and the Simons Foundation for partial support during the completion of this work.

\appendix

\section{Stress-energy tensor in Einstein-Axion-Dilaton Theories}
\label{app:axiondilaton}

Here we wish to demonstrate that the condition \eqref{eq:stresstensorcondition} that we needed to claim we could ignore the matter terms in \eqref{eq:equationwithmatter} holds in a large class of matter theories. We will consider a general Einstein-Axion-Dilaton matter theory with an action of the form  
\begin{equation}
S=\int d^{d+2} x\sqrt{-g}\left(\mathcal{R}-2\Lambda+\mathcal{L}_M\right),
\end{equation}
with a matter Lagrangian
\begin{equation}\label{eq:lagrangian}
\mathcal{L}_M=-\frac{Z(\phi)}{4}F_{\mu\nu}F^{\mu\nu}-\frac{1}{2}\left(\partial\phi\right)^2+V(\phi)-\frac{Y(\phi)}{2}\sum_{i=1}^{d}\left(\partial\varphi_i\right)^2,
\end{equation}
which has an equilibrium black hole solution \eqref{eq:backgroundef} sourced by the fields
\begin{align}
\label{matter}
\phi\equiv\phi(r),&& \varphi_i=mx^i,&& A=A_v(r)dv,
\end{align}
with other components of the Maxwell field $A_\mu$ vanishing.

We wish to demonstrate that for this general class of matter fields \eqref{eq:stresstensorcondition} holds identically for any $\omega, k$. That is
\begin{equation}
\bigg[ T_{vr}(r_0) \delta g_{vv}^{(0)} - \delta T_{vv}(r_0) \bigg] = 0 .
\end{equation}
 For the Lagrangian \eqref{eq:lagrangian} the stress-energy tensor is
\begin{equation}
\begin{aligned}
\label{eq:explicitstresstensor}
T_{\mu\nu}&=\frac{1}{2}\mathcal{L}_M g_{\mu\nu}-\frac{\partial\mathcal{L}_M}{\partial g^{\mu\nu}}\\
&=\frac{1}{2}\mathcal{L}_M g_{\mu\nu}+\frac{1}{2}\partial_\mu\phi\partial_\nu\phi+\frac{Y(\phi)}{2}\sum_{i=1}^{d}\partial_\mu\varphi_i\partial_\nu\varphi_i-\frac{Z(\phi)}{2}F_{\mu\alpha}g^{\alpha\beta}F_{\beta\nu}\,,
\end{aligned}
\end{equation}
from which we can read off the relevant component $T_{vr}$ of the background stress-energy tensor as
\begin{equation}
\label{eq:trv}
T_{vr}=\frac{1}{2}\mathcal{L}_M- \frac{Z(\phi)}{2}F_{vr}^2.
\end{equation}

We also need to work out $\delta T_{vv}$, and hence need to vary the stress-energy tensor \eqref{eq:explicitstresstensor} once more with respect to the fields $g_{\mu\nu},A_{\mu},\phi,\varphi$. At leading order this gives 
\begin{equation}
\begin{aligned}
&\delta T_{\mu\nu}=\frac{1}{2}\mathcal{L}_M\delta g_{\mu\nu}+\frac{Z(\phi)}{2}F_{\mu\alpha}g^{\alpha\gamma}g^{\delta\beta}F_{\beta\nu}\delta g_{\gamma\delta}+\frac{1}{2}g_{\mu\nu}\left[\frac{\delta \mathcal{L}_M}{\delta g^{\alpha\beta}}\delta g^{\alpha\beta}+\frac{\delta \mathcal{L}_M}{\delta\psi_i}\delta\psi_i\right]\\
&+\frac{1}{2}\left[\partial_\mu\phi\partial_\nu\delta\phi+\partial_\mu\delta\phi\partial_\nu\phi+Y(\phi)\sum_{i=1}^{d}\left(\partial_\mu\varphi_i\partial_\nu\delta\varphi_i+\partial_\nu\varphi_i\partial_\mu\delta\varphi_i\right)+\delta\phi\frac{\partial Y(\phi)}{\partial\phi}\sum_{i=1}^{d}\partial_\mu\varphi_i\partial_\nu\varphi_i\right]\\
&-\frac{Z(\phi)}{2}g^{\alpha\beta}\left[F_{\mu\alpha}\left(\partial_\beta\delta A_\nu-\partial_\nu\delta A_\beta\right)+F_{\nu\alpha}\left(\partial_\beta\delta A_\mu-\partial_\mu \delta A_\beta\right)\right]-\frac{Z'(\phi)}{2}F_{\mu\alpha}g^{\alpha\beta}F_{\beta\nu}\delta\phi,
\end{aligned}
\end{equation}
where $\psi_i=\{A_\mu,\phi,\varphi_i\}$ denotes the matter fields.
\\
\\
Evaluating the $vv$-component of this for the black hole solution with the matter fields in \eqref{matter} gives
\begin{eqnarray}
\delta T_{vv} &=& \left(\frac{1}{2}\mathcal{L}_M-\frac{Z(\phi)}{2}F_{vr}^2\right)\delta g_{vv}
+ g_{vv}\bigg[-\frac{Z(\phi)}{2}g_{vv}F_{vr}^2\delta g_{rr}+ Z(\phi) F_{vr}^2 \delta g_{vr}  \nonumber \\ &+& \frac{1}{2}\frac{\delta \mathcal{L}_M}{\delta g^{\alpha\beta}}\delta g^{\alpha\beta} + \frac{1}{2}\frac{\delta \mathcal{L}_M}{\delta\psi_i}\delta\psi_i+Z(\phi)F_{vr}\left(\partial_r\delta A_v-\partial_v\delta A_r\right)-\frac{Z'(\phi)}{2}F_{vr}^2\delta\phi\bigg],
\end{eqnarray}
where we have used that $g^{rv}=1, g^{rr}=-g_{vv}$. 

Now since $g_{vv}(r_0) = -r_0^2f(r_0) = 0$ then assuming the quantity in square brackets is regular at the horizon we immediately see by comparing to \eqref{eq:trv} that
\begin{equation}
\delta T_{vv}(r_0)-T_{vr}(r_0)\delta g_{vv}(r_0)=0,
\end{equation}
indeed holds for this class of black holes. In all these theories the $vv$ component of the Einstein equations therefore reduces at the horizon to the universal form \eqref{perteqn0} that depends only on the near-horizon expansion of the metric perturbations.

\section{Generalisation to AdS$_5$}
\label{app:ads5}

In the main text we studied the AdS$_4$ axion model. However, as we have noted, the phenomenon of pole-skipping was first noticed in a gravitational setting in \cite{Grozdanov:2017ajz}, which studied Einstein gravity in AdS$_5$. Here we will show that the matching argument presented in Section \ref{sec:greensfunction} can easily be generalised to explain the pole-skipping observed in \cite{Grozdanov:2017ajz}. The action we consider is
\begin{equation}
S=\int d^5x\sqrt{-g}\left(\mathcal{R}+12\right),
\end{equation}
and the AdS$_5$-Schwarzschild solution of interest to us can be written
\begin{equation}
ds^2=-r^2f(r)dv^2+2dvdr+r^2\left(dx^2+dy^2+dz^2\right),\quad\quad\quad\quad f(r)=1-\frac{r_0^4}{r^4}.
\end{equation}
in ingoing EF coordinates. Note that for this theory the special location \eqref{location} in Fourier space corresponds to
\begin{align}
\label{eq:locationexplicitads5}
\omega = i  \lambda\,, && \lambda = 2 \pi T= \frac{r_0^2 f'(r_0)}{2}  \,,&&  k = i k_0 \,, && k_0^2 =  \frac{3r_0^3 f'(r_0)}{2} \,.
\end{align}

The generalisation of the `master field'  \eqref{eq:psidef} to Schwarzschild-AdS$_5$ is
\begin{equation}
\begin{aligned}
\label{eq:psidefads5}
\psi=&\;r^4f\Biggl[\frac{d}{dr}\left(\frac{\delta g_{xx}+\delta g_{yy}+\delta g_{zz}}{r^2}\right)-\frac{i\omega}{r^4f}\left(\delta g_{xx}+\delta g_{yy}+\delta g_{zz}\right)-\frac{2ik}{r^2}\left(\delta g_{xr}+\frac{\delta g_{vx}}{r^2f}\right)\\
&\;-3rf\left(\delta g_{rr}+\frac{2}{r^2f}\delta g_{vr}+\frac{1}{r^4f^2}\delta g_{vv}\right)-\frac{\left(k^2+\frac{3}{2}r^3f'\right)}{2r^5f}\left(\delta g_{yy}+\delta g_{zz}\right)\Biggr],
\end{aligned}
\end{equation}
which obeys the equation of motion
\begin{equation}
\frac{d}{dr}\left[\frac{r^3f}{\left(k^2+\frac{3}{2}r^3f'\right)^2}\psi'\right]-\frac{2i\omega r}{\left(k^2+\frac{3}{2}r^3f'\right)^2}\psi'+\frac{\Omega(\omega,k,r)}{r\left(k^2+\frac{3}{2}r^3f'\right)^3}\psi=0,
\end{equation}
where
\begin{equation}
\Omega(\omega,k,r)=-k^4+k^2r\left(6r-i\omega\right)+6r^3\left(4r-5i\omega\right)+f\left[\frac{3}{2}r^5f'-5k^2r^2+6r^3\left(5i\omega-4r\right)\right].
\end{equation}
We again find that when $k^2=-k_0^2$ the two solutions near the horizon are of the form \eqref{onn01} and thus at the special location \eqref{location} the general solution for $\psi$ that is regular at the horizon is of the form \eqref{onn1}.

Infinitesimally away from the location \eqref{location}, the solution continues to take the form \eqref{onn1} away from the horizon with the ratio $a_2/a_1$ fixed by the direction in Fourier space in which one moves. To quantify this, we again parameterise the perturbation away from \eqref{location} by \eqref{eq:scaling1} and note that near the horizon
\begin{equation}
k^2+\frac{3}{2}r^3f'(r)=\epsilon-\tilde{b}(r-r_0)+O((r-r_0)^2),\quad\quad \tilde{b}=12r_0.
\end{equation}

We first solve in the IR regime $r-r_0\sim\epsilon/\tilde{b}$ of the spacetime by scaling the radial coordinate as in \eqref{eq:scaling2} and then solving perturbatively for $\psi$, as in Section \ref{sec:analyticgreensfunction}. After demanding regularity at the horizon, the solution is
\begin{equation}
\label{eq:ads5match}
\psi_r(r)=1+b_+(q)\tilde{b}(r-r_0)+\ldots,
\end{equation}
where $\ldots$ indicate terms that vanish as $\epsilon\rightarrow0$, and
\begin{equation}
b_+(q)=q-\frac{5}{24r_0^2}.
\end{equation}
In the UV region $r-r_0\gg\epsilon/\tilde{b}$, the solution has the form \eqref{onn1}. Matching \eqref{onn1} to \eqref{eq:ads5match} where the solutions overlap tells us that imposing ingoing boundary conditions enforces the relation
\begin{equation}
\label{eq:ads5matching}
\frac{a_2}{a_1}=b_+(q)\tilde{b},
\end{equation}
on the UV solution. In other words the linear combination of $\eta_1$ and $\eta_2$ depends on $q$, the direction in which one moves away from the point \eqref{location} in Fourier space. Unlike in the AdS$_4$ axion case studied in the main text we do not have analytic expressions for $\eta_1$ and $\eta_2$. In this case we therefore do not know which ratio $a_2/a_1$ corresponds to the normalisable mode in the UV and so we cannot analytically predict the slope $q_p$ with which the line of poles passes through the point \eqref{location}.

The analysis we have just performed is consistent with the general argument for pole skipping presented in Section \ref{sec:generalargument}. Using the explicit expression \eqref{eq:psidefads5} for $\psi$ in terms of the metric components, in addition to the non-trivial Einstein equations $X=0$ at \eqref{location}, the relation \eqref{eq:ads5matching} is equivalent to the condition
\begin{equation}
q=\frac{1}{4r_0}\frac{\delta g_{vv}^{(0)}}{\frac{3}{2}h'(r_0)\delta g_{vv}^{(0)}-2\pi T\delta g_{x^ix^i}^{(0)}-2k_0\delta g_{vx}^{(0)}},
\end{equation}
where $\delta g_{\mu\nu}^{(i)}$ are again the coefficients in the near-horizon expansion of the metric components. Converting $q$ back to $\delta\omega/\delta k$ then gives perfect agreement with the equation \eqref{slope1} that relates these coefficients to the direction $q$ in which one moves away from the location \eqref{location}.

\section{Explicit matching to normalisable mode}
\label{app:matching}

In Section~\ref{sec:analyticgreensfunction} we determined the form of the ingoing solution for $\psi$ \eqref{irsolution} near \eqref{location}. Here we will extract the form of the Green's function \eqref{eq:retardedgreenspsi} and a prediction for the slope of the line of poles passing through \eqref{location} by explicitly matching to the UV solution \eqref{eq:uvsol}. In particular we saw in \eqref{uvsol} that after matching the ingoing solution to the UV we are left with a solution of the form 
\begin{equation}
\label{uvsolapp}
\psi_{\mathrm{UV}}(r)  =  1 + \frac{b_+(q) b }{F(r_0)} \int_{r_0}^{r} \mathrm{d}r \frac{\mathrm{exp}\bigg(\frac{m^2 - 3 r_0^2}{r_0 \sqrt{3r_0^2 - 2 m^2}} \mathrm{tan^{-1}}{\frac{(2 r + r_0)}{\sqrt{ 3 r_0^2 - 2 m^2}} \bigg)}}{r \sqrt{2(r^2 + r r_0 + r_0^2)- m^2 }},
\end{equation}
from which the Green's function near \eqref{location} can be extracted by expanding \eqref{uvsolapp} near the UV boundary as $\psi_{\mathrm{UV}} = \psi^{(0)} + \psi^{(1)}/r + \dots $ and 

\begin{equation}
\label{eq:retardedgreenspsiapp}
G^R_{T_{00} T_{00} }(q)= k_0^2(k_0^2 - m^2) \frac{\psi^{(0)}(q)}{\psi^{(1)}(q) - 2 \pi T \psi^{(0)}(q)} + \dots ,
\end{equation}
with $k_0^2$ given by (\ref{locationexplicit}) and the $\dots$ in \eqref{eq:retardedgreenspsiapp} refer to terms that vanish as $\ep \to 0$.\footnote{We will again assume we are not at the special value $m^2 = 2 r_0^2$ so that $k_0^2 - m^2 \neq 0$.}
From expanding \eqref{uvsolapp} one can explicitly read off 
\begin{eqnarray}
\label{uvexp}
\psi^{(0)}(q) = 1 &+& \frac{b_+(q) b N(m,r_0)}{F(r_0)},   \;\;\;\;\ \psi^{(1)}(q) = - \frac{b_+(q) b}{\sqrt{2} F(r_0) } \mathrm{exp}\bigg(\mathrm{sgn}(3 r_0^2 - 2 m^2) \frac{ \pi(m^2 - 3 r_0^2)}{2 r_0 \sqrt{3r_0^2 - 2 m^2}}  \bigg), \nonumber \\
\end{eqnarray}
where have defined the integral
\begin{equation}
\label{N}
N(m,r_0) = \int_{r_0}^{\infty} \mathrm{d}r \frac{\mathrm{exp}\bigg(\frac{m^2 - 3 r_0^2}{r_0 \sqrt{3r_0^2 - 2 m^2}} \mathrm{tan^{-1}}{\frac{(2 r + r_0)}{\sqrt{ 3 r_0^2 - 2 m^2}} \bigg)}}{r \sqrt{2(r^2 + r r_0 + r_0^2)- m^2 }},
\end{equation}
and $F(r_0)$ was defined in \eqref{eq:uvsolexp}. 

We will see a pole in \eqref{eq:retardedgreenspsiapp} when $\eqref{uvsolapp}$ corresponds to a normalisable mode with $2 \pi T \psi^{(0)} = \psi^{(1)}$. From \eqref{uvexp} this implies we need to chose $q = q_p$ by imposing
\begin{equation}
\label{predictionapp}
b b_+(q_p) = - \frac{F(r_0)}{\tilde{N}(m, r_0) },
\end{equation}
with
\begin{equation}
\label{eq:ntilde}
\tilde{N}(m,r_0) =  N(m,r_0) + \frac{1}{\sqrt{2} 2 \pi T} \mathrm{exp}\bigg(\mathrm{sgn}(3 r_0^2 - 2 m^2)\frac{ \pi(m^2 - 3 r_0^2)}{2 r_0 \sqrt{3r_0^2 - 2 m^2}}\bigg)  ,
\end{equation}
which is equivalent to our prediction \eqref{prediction}. Using the explicit forms of $b, T, b_+(q)$ we can use \eqref{predictionapp} to solve for $q_p$ and deduce there is a line of poles passing through \eqref{location} with a slope
\begin{equation}
\frac{1}{v_B} \frac{\delta \omega}{\delta k} = \frac{8(3 r_0^2 - m^2)}{3(2 r_0^2 - m^2) } + \frac{4 (6 r_0^2 - m^2 )}{3(m^2 - 2 r_0^2)} \frac{F(r_0) r_0}{{\tilde N}(m, r_0)  }.
\end{equation} 

Likewise there will be a line of zeroes when $\eqref{uvsolapp}$ corresponds to the non-normalisable solution with $\psi^{(0)} = 0$. From \eqref{uvexp} this corresponds to choosing $q = q_z$ such that
\begin{equation}
b b_+(q_z) = -\frac{F(r_0)}{ N(m, r_0) }.
\end{equation}
Indeed from \eqref{uvsolapp} it is straightforward to read off the Green's function itself as 
\begin{equation}
G^R_{T_{00} T_{00} }(q) = -\frac{k_0^2(k_0^2-m^2)}{2 \pi T} \frac{F(r_0)  + b b_+(q) N(m,r_0)}{F(r_0)  + b b_+(q) \tilde{N}(m,r_0)},
\end{equation}
from which one can explicitly see the line of poles corresponding to moving away from \eqref{location} along the slope $q= q_p$ and a line of zeroes for $q = q_z$.

\section{Details of calculations at the $SL(2,R) \times SL(2,R)$ invariant point}
\label{app:SL2}

Here we wish to discuss in detail the special case of $m^2 = 2 r_0^2$, in order to both highlight various subtleties with the discussion in terms of the gauge invariant mode $\psi$ and also to use this case to elaborate on the general argument for pole-skipping we provided in Section~\ref{sec:generalargument}.

\subsection*{Pole-skipping in terms of $\psi$}

In particular, whilst our arguments in Sections~\ref{sec:specialpoint} and~\ref{sec:analyticgreensfunction} are generically valid, the description of pole skipping in terms of the variable $\psi$ is somewhat different for the choice of $m^2 = 2 r_0^2$.  To see why $m^2 = 2 r_0^2$ is special it is useful to note that the function $k^2 + r^3 f'(r) = k^2 + m^2$ that appears in the equation \eqref{eq:psieom} for $\psi$ becomes a constant for this value of $m$. At the special point \eqref{location} $k^2 = - k_0^2 = - m^2$ it therefore vanishes identically, and we must be more careful in describing the behaviour of \eqref{eq:psieom} at \eqref{location}. For the choice $m^2 = 2 r_0^2$ the equation of motion \eqref{eq:psieom} dramatically simplifies to
\begin{equation}
\label{eq:specialeom}
\frac{d}{d r} (r^2 f \psi') - 2 i \omega \psi' - \frac{k^2}{r^2} \psi = 0,
\end{equation}
from which the retarded energy density Green's function can be extracted by solving \eqref{eq:specialeom} subject to ingoing boundary conditions and using
\begin{equation}
\label{eq:retardedgreenspsi2}
G^R_{T_{00} T_{00} }(\omega,k)= k^2(k^2 + 2 r_0^2) \frac{\psi^{(0)}(\omega,k)}{\psi^{(1)}(\omega,k) + i \omega \psi^{(0)}(\omega,k)}. 
\end{equation}
Note that upon setting $\omega = i \lambda$ and $k = i k_0$ in \eqref{eq:specialeom} this no longer reduces to the equation \eqref{eq:psieom22} which holds for any other $m^2 \neq 2 r_0^2$. Solving \eqref{eq:specialeom} at \eqref{location} one now finds that the two linearly independent solutions for $\psi$ are
\begin{equation}
\label{specialsolutions}
\psi_1 =  c_1 \frac{r + r_0}{r}, \;\;\;\;\; \psi_2 =   \frac{c_2}{ r (r - r_0)} \bigg[2 r r_0 + (r_0^2 - r^2)\mathrm{log}\bigg(\frac{r-r_0}{r + r_0} \bigg) \bigg],
\end{equation}
from which we see that there is only one solution $\psi_1$ that is regular at the horizon. 

Therefore whilst for any other $m^2 \neq 2 r_0^2$ there are two linearly independent regular solutions for $\psi$ at \eqref{location}, for the specific choice $m^2  = 2 r_0^2$ there is instead a unique regular solution. Note that this observation is perfectly consistent with smoothly taking the limit $m^2 \to 2 r_0^2$ in the solution \eqref{uvsol} we derived by matching, even though the matching procedure is formally invalid at this point as $b\rightarrow0$. In that limit we find $b b_+(q) = -1/2 r_0$ and hence the ingoing solution in \eqref{uvsol} becomes independent of the slope. Upon performing the integral in \eqref{uvsol} for $m^2 = 2 r_0^2$ one finds precisely that the solution becomes $\psi_{\mathrm{UV}} = \psi_1$.\footnote{Furthermore, this also is consistent our discussion in Section~\ref{sec:generalargument} that there should be an extra free parameter for the metric solutions at \eqref{location} corresponding to $\delta g_{vv}^{(0)}$. One can see directly from \eqref{aplus} that we should still only find a unique ingoing mode for $\psi$ at $b=0$ even when the metric component $\delta g_{vv}^{(0)}$ is a free parameter.}

Note that although there is a unique solution for $\psi_1$, the Green's function \eqref{eq:retardedgreenspsi2} still cannot be defined at this point. To see this note that the ingoing solution $\psi_1$ is normalisable in the UV, that is the denominator in \eqref{eq:retardedgreenspsi2} vanishes for $\psi_1$. However the numerator in \eqref{eq:retardedgreenspsi2} vanishes at $k^2 = - k_0^2$ because of the explicit factor of $k^2 + 2 r_0^2$ in the relation between $G^R_{T_{00} T_{00} }(\omega,k)$ and the asymptotics of $\psi$. 

{In contrast to our explanation in Section~\ref{sec:generalargument}, at this value of $m$ it appears that the pole-skipping phenomenon is unrelated to the existence of an extra ingoing solution. As we will illustrate shortly, this is not the case. The aforementioned subtleties associated with $\psi$ at the point $m^2=2r_0^2,\omega=i\lambda, k=ik_0$ are in fact indicators that $\psi$ is not a suitable variable for capturing the general gauge-invariant solution to the Einstein equations at this point. A more careful analysis reveals that at this point $\psi$ in fact obeys a first order equation of motion (which enforces $c_2=0$), as does the other decoupled gauge-invariant variable (see \eqref{eq:GIvariablesschwarz} below). To illustrate the origin of pole-skipping at this value of $m$, we will therefore examine the solutions for the metric perturbations directly.}

\subsection*{Discussion in terms of metric solutions} 

The enhanced symmetry at $m^2=2r_0^2$ allows the linearised gravitational equations of motion to be solved analytically. The following is an exact solution to these equations at $m=2r_0^2,\omega=i\lambda,k=ik_0$ in ingoing EF coordinates
\begin{equation*}
\begin{aligned}
\delta g_{vv}=&\frac{e^{r_0(v-\sqrt{2}x)}}{8r^3}\Biggl[r_0^2\left(\sqrt{2}\delta g_{tx}^{(0)}+\delta g_{yy}^{(0)}-\delta\varphi_1^{(0)}\right)\left(4r^3+4r^2r_0+rr_0^2-r_0^3\right)+16cr(r+r_0)\\
&+8\delta g_{tt}^{(0)}\left(r^5+r^4r_0-r_0^5\right)\Biggr],
\end{aligned}
\end{equation*}
\begin{equation*}
\begin{aligned}
\delta g_{vx}=&\,\frac{e^{r_0(v-\sqrt{2}x)}}{40r^3}\Biggl[5(r+r_0)\left(\left(8r^4-4r^2r_0^2-r_0^4\right)\delta g_{tx}^{(0)}-2\sqrt{2}r_0^2\left(2r^2+r_0^2\right)\left(\delta g_{yy}^{(0)}-\delta\varphi_1^{(0)}\right)\right)\\
&-8\sqrt{2}c\left(5r^2+5rr_0+3r_0^2\right)\Biggr],
\end{aligned}
\end{equation*}
\begin{equation*}
\begin{aligned}
\delta g_{xx}=&\,\frac{e^{r_0(v-\sqrt{2}x)}}{60r^3}\Biggl[60\delta g_{xx}^{(0)}r^5+60\delta g_{xx}^{(0)}r^4r_0+16c\left(5r^2+5rr_0+6r_0^2\right)+3r_0^5\Bigl(18\delta g_{tt}^{(0)}+3\sqrt{2}\delta g_{tx}^{(0)}\\
&+\delta g_{xx}^{(0)}+17\delta g_{yy}^{(0)}-16\delta\varphi_1^{(0)}\Bigr)-10r^2r_0^3\Bigl(\delta g_{tt}^{(0)}+\sqrt{2}\delta g_{tx}^{(0)}-\delta g_{xx}^{(0)}-4\delta g_{yy}^{(0)}+3\delta\varphi_1^{(0)}\Bigr)\\
&-30r^3r_0^2\Bigl(\delta g_{tt}^{(0)}+\sqrt{2}\delta g_{tx}^{(0)}+\delta g_{xx}^{(0)}-2\delta g_{yy}^{(0)}+3\delta\varphi_1^{(0)}\Bigr)\\
&-5rr_0^4\Bigl(2\delta g_{tt}^{(0)}+2\sqrt{2}\delta g_{tx}^{(0)}+\delta g_{xx}^{(0)}-8\delta g_{yy}^{(0)}+9\delta\varphi_1^{(0)}\Bigr)\Biggr],
\end{aligned}
\end{equation*}
\begin{equation*}
\begin{aligned}
\delta g_{yy}=&\, \frac{e^{r_0(v-\sqrt{2}x)}}{60r^3}\Biggl[60\delta g_{yy}^{(0)}r^5+60\delta g_{yy}^{(0)}r^4r_0+8c\left(5r^2+5rr_0+6r_0^2\right)+3r_0^5\Bigl(22\delta g_{tt}^{(0)}+7\sqrt{2}\delta g_{tx}^{(0)}\\
&-\delta g_{xx}^{(0)}+8\delta g_{yy}^{(0)}-9\delta\varphi_1^{(0)}\Bigr)+10r^2r_0^3\Bigl(\delta g_{tt}^{(0)}+\sqrt{2}\delta g_{tx}^{(0)}-\delta g_{xx}^{(0)}+2\delta g_{yy}^{(0)}-3\delta\varphi_1^{(0)}\Bigr)\\
&+30r^3r_0^2\Bigl(\delta g_{tt}^{(0)}+\sqrt{2}\delta g_{tx}^{(0)}+\delta g_{xx}^{(0)}+\delta\varphi_1^{(0)}\Bigr)+5rr_0^4\Bigl(2\delta g_{tt}^{(0)}+2\sqrt{2}\delta g_{tx}^{(0)}+\delta g_{xx}^{(0)}+\delta g_{yy}^{(0)}\Bigr)\Biggr],
\end{aligned}
\end{equation*}
\begin{equation*}
\begin{aligned}
\delta g_{vr}=&-\frac{e^{r_0(v-\sqrt{2}x)}}{80r^4\left(r+r_0\right)}\Biggl[80\delta g_{tt}^{(0)}r^5+160\delta g_{tt}^{(0)}r^4r_0+32c\left(5r^2+10rr_0+4r_0^2\right)+r_0^5\Bigl(56\delta g_{tt}^{(0)}\\
&+41\sqrt{2}\delta g_{tx}^{(0)}-8\delta g_{xx}^{(0)}+49\delta g_{yy}^{(0)}-57\delta\varphi_1^{(0)}\Bigr)+80r^2r_0^3\Bigl(2\delta g_{tt}^{(0)}+\sqrt{2}\delta g_{tx}^{(0)}+\delta g_{yy}^{(0)}-\delta\varphi_1^{(0)}\Bigr)\\
&+40r^3r_0^2\Bigl(4\delta g_{tt}^{(0)}+\sqrt{2}\delta g_{tx}^{(0)}+\delta g_{yy}^{(0)}-\delta\varphi_1^{(0)}\Bigr)\\
&+10rr_0^4\Bigl(16\delta g_{tt}^{(0)}+9\sqrt{2}\delta g_{tx}^{(0)}+9\delta g_{yy}^{(0)}-9\delta\varphi_1^{(0)}\Bigr)\Biggr],
\end{aligned}
\end{equation*}
\begin{equation*}
\begin{aligned}
\delta g_{xr}=&\,\frac{e^{r_0(v-\sqrt{2}x)}}{40\sqrt{2}r^4\left(r+r_0\right)}\Biggl[16c\left(5r^2+10rr_0+7r_0^2\right)-\sqrt{2}\delta g_{tx}^{(0)}\Bigl(40r^5+80r^4r_0+60r^3r_0^2+40r^2r_0^3\\
&+35rr_0^4+6r_0^5\Bigr)+8r_0^5\Bigl(3\delta g_{tt}^{(0)}+\delta g_{xx}^{(0)}\Bigr)+r_0^2\delta g_{yy}^{(0)}\Bigl(40r^3+80r^2r_0+100rr_0^2+61r_0^3\Bigr)\\
&-r_0^2\delta\varphi_1^{(0)}\Bigl(40r^3+80r^2r_0+100rr_0^2+53r_0^3\Bigr)\Biggr],
\end{aligned}
\end{equation*}
\begin{equation*}
\begin{aligned}
\delta g_{rr}=&\,\frac{e^{r_0(v-\sqrt{2}x)}}{40r^5\left(r+r_0\right)^2}\Biggl[40\delta g_{tt}^{(0)}r^5+120\delta g_{tt}^{(0)}r^4r_0+16c\left(5r^2+15rr_0+8r_0^2\right)+8r_0^5\Bigl(22\delta g_{tt}^{(0)}\\
&+7\sqrt{2}\delta g_{tx}^{(0)}-\delta g_{xx}^{(0)}+8\delta g_{yy}^{(0)}-9\delta\varphi_1^{(0)}\Bigr)+20r^2r_0^3\Bigl(14\delta g_{tt}^{(0)}+3\sqrt{2}\delta g_{tx}^{(0)}+3\delta g_{yy}^{(0)}-3\delta\varphi_1^{(0)}\Bigr)\\
&+20r^3r_0^2\Bigl(10\delta g_{tt}^{(0)}+\sqrt{2}\delta g_{tx}^{(0)}+\delta g_{yy}^{(0)}-\delta\varphi_1^{(0)}\Bigr)\\
&+15rr_0^4\Bigl(24\delta g_{tt}^{(0)}+7\sqrt{2}\delta g_{tx}^{(0)}+7\delta g_{yy}^{(0)}-7\delta\varphi_1^{(0)}\Bigr)\Biggr],
\end{aligned}
\end{equation*}
\begin{equation*}
\begin{aligned}
\delta \varphi_1=&\,\frac{e^{r_0(v-\sqrt{2}x)}}{120r^5}\Biggl[120\delta \varphi_{1}^{(0)}r^5+120\delta\varphi_{1}^{(0)}r^4r_0-8c\left(5r^2+5rr_0+6r_0^2\right)+3r_0^5\Bigl(4\delta g_{tt}^{(0)}+4\sqrt{2}\delta g_{tx}^{(0)}\\
&-2\delta g_{xx}^{(0)}-9\delta g_{yy}^{(0)}+7\delta\varphi_1^{(0)}\Bigr)+20r^2r_0^3\Bigl(\delta g_{tt}^{(0)}+\sqrt{2}\delta g_{tx}^{(0)}-\delta g_{xx}^{(0)}-\delta g_{yy}^{(0)}\Bigr)\\
&+60r^3r_0^2\Bigl(\delta g_{tt}^{(0)}+\sqrt{2}\delta g_{tx}^{(0)}+\delta g_{xx}^{(0)}-\delta g_{yy}^{(0)}+2\delta\varphi_1^{(0)}\Bigr)\\
&+5rr_0^4\Bigl(4\delta g_{tt}^{(0)}+4\sqrt{2}\delta g_{tx}^{(0)}+2\delta g_{xx}^{(0)}-7\delta g_{yy}^{(0)}+9\delta\varphi_1^{(0)}\Bigr)\Biggr].
\end{aligned}
\end{equation*}
We are working here in a different gauge than in Section~\ref{sec:generalargument}. This solution is regular at the horizon $r=r_0$ and depends on six constants: the five sources of the dual energy-momentum tensor and scalar operator $\left\{\delta g_{tt}^{(0)},\delta g_{tx}^{(0)},\delta g_{xx}^{(0)},\delta g_{yy}^{(0)},\delta \varphi_{1}^{(0)}\right\}$, and an additional constant $c$. Consistently with our general argument in Section~\ref{sec:generalargument}, it is manifest that the dependence on $c$ means that fixing the sources does not uniquely specify the ingoing solution. 

Furthermore, it is straightforward to explicitly check that the dependence on the additional constant $c$ results in the retarded Green's function of energy density being infinitely multi-valued. Upon transforming to the coordinates \eqref{eq:background} and expanding near the boundary $r\rightarrow\infty$, the solution is
\begin{equation}
\begin{aligned}
&\delta g_{tt}=e^{r_0(t-\sqrt{2}x)}\left[\delta g_{tt}^{(0)}r^2+\ldots+\frac{2c}{r}+\ldots\right]\!,\quad\quad \delta g_{tx}=e^{r_0(t-\sqrt{2}x)}\left[\delta g_{tx}^{(0)}r^2+\ldots-\frac{\sqrt{2}c}{r}+\ldots\right]\!,\\
&\delta g_{xx}=e^{r_0(t-\sqrt{2}x)}\!  \left[\delta g_{xx}^{(0)}r^2+\ldots+\frac{1}{3r}\left[4c+r_0^3\left(\delta g_{tt}^{(0)}+\sqrt{2}\delta g_{tx}^{(0)}+2\delta g_{xx}^{(0)}-\delta g_{yy}^{(0)}+3\delta\varphi^{(0)}_1\right)\right]+\ldots\right]\!,\\
&\delta g_{yy}=e^{r_0(t-\sqrt{2}x)}\! \left[\delta g_{yy}^{(0)}r^2+\ldots+\frac{1}{3r}\left[2c-r_0^3\left(\delta g_{tt}^{(0)}+\sqrt{2}\delta g_{tx}^{(0)}+2\delta g_{xx}^{(0)}-\delta g_{yy}^{(0)}+3\delta\varphi^{(0)}_1\right)\right]+\ldots\right]\!,\\
&\delta \varphi_1=e^{r_0(t-\sqrt{2}x)}\left[\delta\varphi^{(0)}_1+\ldots-\frac{1}{3r^3}\left[c+r_0^3\left(\delta g_{tt}^{(0)}+\sqrt{2}\delta g_{tx}^{(0)}+2\delta g_{xx}^{(0)}-\delta g_{yy}^{(0)}+3\delta\varphi^{(0)}_1\right)\right]+\ldots\right]\!,\\
&\delta g_{tr}=e^{r_0(t-\sqrt{2}x)}O(r^{-5}),\quad\quad\delta g_{xr}=e^{r_0(t-\sqrt{2}x)}O(r^{-5}),\quad\quad\delta g_{rr}=e^{r_0(t-\sqrt{2}x)}O(r^{-7}).\\
\end{aligned}
\end{equation}
Calculating the expectation value of the dual energy-momentum tensor of this solution using \cite{Bianchi:2001kw,Andrade:2013gsa}
\begin{equation}
\langle T^{\mu\nu}\rangle=\lim_{r\rightarrow\infty}r^5\left[2\left(K^{\mu\nu}-K\gamma^{\mu\nu}+G_{\gamma}^{\mu\nu}-2\gamma^{\mu\nu}\right)+\frac{1}{2}\gamma^{\mu\nu}\partial\varphi_i\cdot\partial\varphi_i-\nabla^\mu\varphi_i\nabla^\nu\varphi_i\right],
\end{equation}
yields
\begin{equation}
\begin{aligned}
\label{eq:exactEMtensor}
&\langle T^{tt} \rangle = e^{r_0(t-\sqrt{2}x)}6c,\\
&\langle T^{tx} \rangle = e^{r_0(t-\sqrt{2}x)}3\sqrt{2}c,\\
&\langle T^{xx} \rangle = e^{r_0(t-\sqrt{2}x)}\left[4c+r_0^3\left(\delta g_{tt}^{(0)}+\sqrt{2}\delta g_{tx}^{(0)}+2\delta g_{xx}^{(0)}-\delta g_{yy}^{(0)}+3\delta\varphi_1^{(0)}\right)\right],\\
&\langle T^{yy} \rangle = e^{r_0(t-\sqrt{2}x)}\left[2c-r_0^3\left(\delta g_{tt}^{(0)}+\sqrt{2}\delta g_{tx}^{(0)}+2\delta g_{xx}^{(0)}-\delta g_{yy}^{(0)}+3\delta\varphi_1^{(0)}\right)\right].
\end{aligned}
\end{equation}
The expectation value of the energy density $\langle T^{tt}\rangle$ depends on the arbitrary constant $c$ and thus the corresponding Green's function is not well-defined (consistent with the general expression \eqref{eq:exactcorrelator}). Our analysis here makes it clear that this property can be traced back to the presence of an additional ingoing solution to the equations of motion.

\section{Details of numerical calculations}
\label{app:numerics}

The full dispersion relations $\omega(k)$ of the poles of $G^R_{T^{00}T^{00}}$ can be found by solving Einstein's equations numerically. In this appendix we briefly describe how we did this to produce the numerical results shown in Figures \ref{fig:plots} and \ref{fig:plot} of section \ref{sec:hydropoles}.

To perform the numerical calculations, we transformed to a coordinate system in which the metric is
\begin{equation}
ds^2=-r^2f(r)dt^2+r^2\left(dx^2+dy^2\right)+\frac{dr^2}{r^2f(r)}.
\end{equation}
We studied linear perturbations of the metric $\delta \tilde{g}_{\mu\nu}(r,t,x)$ and the scalar fields $\delta\tilde{\varphi}_i(r,t,x)$ around this spacetime, after Fourier transforming in the $(t,x)$ coordinates to the conjugate variables $(\tilde{\omega},k)$ (the tildes differentiate quantities from the analogous ones in ingoing EF coordinates studied in the main text). The perturbation of interest to us is $\delta\tilde{g}_{tt}$, and it couples to $\delta\tilde{g}_{xt}$, $\delta\tilde{g}_{xx}$, $\delta\tilde{g}_{yy}$, $\delta\tilde{g}_{rr}$, $\delta\tilde{g}_{xr}$, $\delta\tilde{g}_{tr}$ and $\delta\tilde{\varphi}_{1}$. In addition to the three first-order (in radial derivatives) equations governing these fields, there are two second-order equations. The Green's function $G^R_{T^{00}T^{00}}$ can be efficiently extracted by writing the two second-order equations as a closed set of equations for two gauge-invariant variables \cite{Kovtun:2005ev}. A convenient choice for these variables is
\begin{equation}
\begin{aligned}
\label{eq:GIvariablesschwarz}
\Psi_1=&\;\frac{r^2f}{\left(k^2+r^3f'\right)^2}\frac{d}{dr}\Biggl[r^4f\left(\frac{d}{dr}\left(\frac{\delta\tilde{g}_{xx}+\delta\tilde{g}_{yy}}{r^2}\right)-\frac{2ik}{r^2}\delta\tilde{g}_{xr}-2rf\delta\tilde{g}_{rr}-\frac{\left(k^2+r^3f'\right)}{r^5f}\delta\tilde{g}_{yy}\right)\\
&-\frac{\left(k^2+r^3f'\right)}{\left(k^2+m^2\right)}\frac{mr}{2}\left\{m\left(\frac{\delta\tilde{g}_{xx}-\delta\tilde{g}_{yy}}{r^2}\right)-2ik\delta\tilde{\varphi}_1\right\}\Biggr],\\
\Psi_2=&\;m\left(\frac{\delta\tilde{g}_{xx}-\delta\tilde{g}_{yy}}{r^2}\right)-2ik\delta\tilde{\varphi}_1,
\end{aligned}
\end{equation}
as they obey two equations of motion that decouple and thus can be solved independently. The equation for $\Psi_1$ is
\begin{equation}
\label{eq:Psinumericeq}
\frac{d}{dr}\left[\frac{r^2f\left(k^2+r^3f'\right)^3}{\tilde{\omega}^2\left(k^2+r^3f'\right)-k^2\left(k^2+m^2\right)f}\Psi_1'(r)\right]+\frac{\left(k^2+r^3f'\right)^2}{r^2f}\Psi_1(r)=0.
\end{equation}

Our first method for extracting the poles of $G^R_{T^{00}T^{00}}(\omega,k)$ was to solve \eqref{eq:Psinumericeq} numerically by imposing ingoing boundary conditions $\Psi_1\propto(r-r_0)^{-i\tilde{\omega}/4\pi T}$ near the horizon and integrating to the boundary. We then extracted the coefficients of the near-boundary expansion $\Psi_1(r\rightarrow\infty)=\Psi_1^{(0)}+\Psi_1^{(1)}r^{-1}+\ldots$ and looked for zeroes of the quantity $\Psi_1^{(0)}/\Psi_1^{(1)}$, which correspond to poles of $G^R_{T^{00}T^{00}}(\omega,k)$. A drawback of this method is that for imaginary values of $k$ the equation develops a singularity within the integration region when $\left|k^2\right|$ lies between $m^2$ and $\left(2\pi T/v_B\right)^2$. This is an artifact of the variables chosen and means that we cannot track the poles over the entire range of imaginary $k$ that we desire.

To circumvent this problem, we also considered a different variable
\begin{equation}
\Psi_3=\frac{1}{r^2}\left(\tilde{\omega}^2\delta\tilde{g}_{xx}+2\tilde{\omega}k\delta\tilde{g}_{xt}+k^2\delta\tilde{g}_{tt}-\left(\tilde{\omega}^2-k^2f-\frac{1}{2}k^2rf'\right)\delta\tilde{g}_{yy}\right),
\end{equation}
which obeys the equation of motion
\begin{equation}\label{eq:Psi3eqnumeric}
\begin{aligned}
\frac{d}{dr}\left[\frac{r^4f}{\left(\tilde{\omega}^2-k^2f-\frac{k^2}{4}rf'\right)^2-m^2f\left(\tilde{\omega}^2-k^2\left(\frac{3}{4}-\frac{m^2}{8r^2}\right)\right)}\Psi_3'\right]& \\
+\frac{32r^4A_1(r)}{fA_2(r)}\Psi_3+A_3(r)\Psi_2'+A_4(r)\Psi_2 & =0,
\end{aligned}
\end{equation}
where
\begin{equation}
\begin{aligned}
A_1(r)=&\;2\left\{k^2\tilde{\omega}\left(m^2-6r^2\right)+8r^2\tilde{\omega}^3\right\}^2 \\
&-f\left\{k^2\left(m^2-6r^2\right)+8r^2\tilde{\omega}^2\right\}\Bigl[k^2\left(m^2-6r^2\right)\left(2k^2+5m^2-6r^2\right)\\
&+8r^2\tilde{\omega}^2\left(3k^2+4m^2\right)\Bigr]+2r^2f^2\Bigl[k^4\left\{7m^4+36m^2r^2-180r^4+4k^2\left(m^2-6r^2\right)\right\}\\
&+4r^2\tilde{\omega}^2\left(9k^4+16m^4+72k^2r^2\right)\Bigr] \\
&-4k^2r^4f^3\left\{2k^4+k^2\left(11m^2-18r^2\right)+24\left(m^4+3r^2\omega^2\right)\right\}+72k^4r^6f^4,\\
A_2(r)=&\;\Big[\left\{k^2\left(m^2-6r^2\right)+8r^2\tilde{\omega}^2\right\}^2 \\
& -4r^2f\left(k^2+2m^2\right)\left\{k^2\left(m^2-6r^2\right)+8r^2\tilde{\omega}^2\right\}+4k^4r^4f^2\Big]^2,
\end{aligned}
\end{equation}
and the precise forms of $A_3(r)$ and $A_4(r)$ are not important. The solutions that determine the poles of $G^R_{T^{00}T^{00}}(\omega,k)$ are those where the boundary metric is unchanged, which requires that $\Psi_2(r\rightarrow\infty)=0$. As $\Psi_2(r)$ obeys a decoupled, linear equation of motion then the ingoing solution that vanishes at the boundary will generically be $\Psi_2(r)=0$ and so our second method was therefore to numerically integrate the equation \eqref{eq:Psi3eqnumeric} with $\Psi_2(r)=0$, after imposing ingoing boundary conditions on $\Psi_3\propto(r-r_0)^{-i\tilde{\omega}/4\pi T}$ at the horizon. After expanding near the boundary $\Psi_3(r\rightarrow\infty)=\Psi_3^{(0)}+\Psi_3^{(1)}r^{-1}+\ldots$ we then extracted the poles of $G^R_{T^{00}T^{00}}(\omega,k)$ by determining the zeroes of $\Psi_3^{(0)}/\Psi_3^{(3)}$. The exception to $\Psi_2(r)=0$ being the appropriate solution is at particular sets of $(\tilde{\omega},k)$ that correspond to quasinormal modes of the $\Psi_2$ equation, but we expect that generically these frequencies will not overlap with the poles of $G^R_{T^{00}T^{00}}(\omega,k)$. We confirmed that this is the case by checking that our results from this method agree with the results from the previous method in the regime where both could be performed.

\bibliography{references}

\begin{thebibliography}{48}%
\makeatletter
\providecommand \@ifxundefined [1]{%
 \@ifx{#1\undefined}
}%
\providecommand \@ifnum [1]{%
 \ifnum #1\expandafter \@firstoftwo
 \else \expandafter \@secondoftwo
 \fi
}%
\providecommand \@ifx [1]{%
 \ifx #1\expandafter \@firstoftwo
 \else \expandafter \@secondoftwo
 \fi
}%
\providecommand \natexlab [1]{#1}%
\providecommand \enquote  [1]{``#1''}%
\providecommand \bibnamefont  [1]{#1}%
\providecommand \bibfnamefont [1]{#1}%
\providecommand \citenamefont [1]{#1}%
\providecommand \href@noop [0]{\@secondoftwo}%
\providecommand \href [0]{\begingroup \@sanitize@url \@href}%
\providecommand \@href[1]{\@@startlink{#1}\@@href}%
\providecommand \@@href[1]{\endgroup#1\@@endlink}%
\providecommand \@sanitize@url [0]{\catcode `\\12\catcode `\$12\catcode
  `\&12\catcode `\#12\catcode `\^12\catcode `\_12\catcode `\%12\relax}%
\providecommand \@@startlink[1]{}%
\providecommand \@@endlink[0]{}%
\providecommand \url  [0]{\begingroup\@sanitize@url \@url }%
\providecommand \@url [1]{\endgroup\@href {#1}{\urlprefix }}%
\providecommand \urlprefix  [0]{URL }%
\providecommand \Eprint [0]{\href }%
\providecommand \doibase [0]{http://dx.doi.org/}%
\providecommand \selectlanguage [0]{\@gobble}%
\providecommand \bibinfo  [0]{\@secondoftwo}%
\providecommand \bibfield  [0]{\@secondoftwo}%
\providecommand \translation [1]{[#1]}%
\providecommand \BibitemOpen [0]{}%
\providecommand \bibitemStop [0]{}%
\providecommand \bibitemNoStop [0]{.\EOS\space}%
\providecommand \EOS [0]{\spacefactor3000\relax}%
\providecommand \BibitemShut  [1]{\csname bibitem#1\endcsname}%
\let\auto@bib@innerbib\@empty
%</preamble>
\bibitem [{\citenamefont {{Larkin}}\ and\ \citenamefont
  {{Ovchinnikov}}(1969)}]{1969JETP...28.1200L}%
  \BibitemOpen
  \bibfield  {author} {\bibinfo {author} {\bibfnamefont {A.~I.}\ \bibnamefont
  {{Larkin}}}\ and\ \bibinfo {author} {\bibfnamefont {Y.~N.}\ \bibnamefont
  {{Ovchinnikov}}},\ }\bibfield  {title} {\enquote {\bibinfo {title}
  {{Quasiclassical Method in the Theory of Superconductivity}},}\ }\href@noop
  {} {\bibfield  {journal} {\bibinfo  {journal} {Soviet Journal of Experimental
  and Theoretical Physics}\ }\textbf {\bibinfo {volume} {28}},\ \bibinfo
  {pages} {1200} (\bibinfo {year} {1969})}\BibitemShut {NoStop}%
\bibitem [{\citenamefont {Shenker}\ and\ \citenamefont
  {Stanford}(2014)}]{Shenker:2013pqa}%
  \BibitemOpen
  \bibfield  {author} {\bibinfo {author} {\bibfnamefont {Stephen~H.}\
  \bibnamefont {Shenker}}\ and\ \bibinfo {author} {\bibfnamefont {Douglas}\
  \bibnamefont {Stanford}},\ }\bibfield  {title} {\enquote {\bibinfo {title}
  {{Black holes and the butterfly effect}},}\ }\href {\doibase
  10.1007/JHEP03(2014)067} {\bibfield  {journal} {\bibinfo  {journal} {JHEP}\
  }\textbf {\bibinfo {volume} {03}},\ \bibinfo {pages} {067} (\bibinfo {year}
  {2014})},\ \Eprint {http://arxiv.org/abs/1306.0622} {arXiv:1306.0622
  [hep-th]} \BibitemShut {NoStop}%
%%CITATION = ARXIV:1306.0622;%%
\bibitem [{\citenamefont {Roberts}\ \emph {et~al.}(2015)\citenamefont
  {Roberts}, \citenamefont {Stanford},\ and\ \citenamefont
  {Susskind}}]{Roberts:2014isa}%
  \BibitemOpen
  \bibfield  {author} {\bibinfo {author} {\bibfnamefont {Daniel~A.}\
  \bibnamefont {Roberts}}, \bibinfo {author} {\bibfnamefont {Douglas}\
  \bibnamefont {Stanford}}, \ and\ \bibinfo {author} {\bibfnamefont {Leonard}\
  \bibnamefont {Susskind}},\ }\bibfield  {title} {\enquote {\bibinfo {title}
  {{Localized shocks}},}\ }\href {\doibase 10.1007/JHEP03(2015)051} {\bibfield
  {journal} {\bibinfo  {journal} {JHEP}\ }\textbf {\bibinfo {volume} {03}},\
  \bibinfo {pages} {051} (\bibinfo {year} {2015})},\ \Eprint
  {http://arxiv.org/abs/1409.8180} {arXiv:1409.8180 [hep-th]} \BibitemShut
  {NoStop}%
%%CITATION = ARXIV:1409.8180;%%
\bibitem [{\citenamefont {Shenker}\ and\ \citenamefont
  {Stanford}(2015)}]{Shenker:2014cwa}%
  \BibitemOpen
  \bibfield  {author} {\bibinfo {author} {\bibfnamefont {Stephen~H.}\
  \bibnamefont {Shenker}}\ and\ \bibinfo {author} {\bibfnamefont {Douglas}\
  \bibnamefont {Stanford}},\ }\bibfield  {title} {\enquote {\bibinfo {title}
  {{Stringy effects in scrambling}},}\ }\href {\doibase
  10.1007/JHEP05(2015)132} {\bibfield  {journal} {\bibinfo  {journal} {JHEP}\
  }\textbf {\bibinfo {volume} {05}},\ \bibinfo {pages} {132} (\bibinfo {year}
  {2015})},\ \Eprint {http://arxiv.org/abs/1412.6087} {arXiv:1412.6087
  [hep-th]} \BibitemShut {NoStop}%
%%CITATION = ARXIV:1412.6087;%%
\bibitem [{\citenamefont {Maldacena}\ \emph
  {et~al.}(2016{\natexlab{a}})\citenamefont {Maldacena}, \citenamefont
  {Shenker},\ and\ \citenamefont {Stanford}}]{Maldacena:2015waa}%
  \BibitemOpen
  \bibfield  {author} {\bibinfo {author} {\bibfnamefont {Juan}\ \bibnamefont
  {Maldacena}}, \bibinfo {author} {\bibfnamefont {Stephen~H.}\ \bibnamefont
  {Shenker}}, \ and\ \bibinfo {author} {\bibfnamefont {Douglas}\ \bibnamefont
  {Stanford}},\ }\bibfield  {title} {\enquote {\bibinfo {title} {{A bound on
  chaos}},}\ }\href {\doibase 10.1007/JHEP08(2016)106} {\bibfield  {journal}
  {\bibinfo  {journal} {JHEP}\ }\textbf {\bibinfo {volume} {08}},\ \bibinfo
  {pages} {106} (\bibinfo {year} {2016}{\natexlab{a}})},\ \Eprint
  {http://arxiv.org/abs/1503.01409} {arXiv:1503.01409 [hep-th]} \BibitemShut
  {NoStop}%
%%CITATION = ARXIV:1503.01409;%%
\bibitem [{\citenamefont {Kitaev}(2014)}]{kitaev}%
  \BibitemOpen
  \bibfield  {author} {\bibinfo {author} {\bibfnamefont {A.}~\bibnamefont
  {Kitaev}},\ }\bibfield  {title} {\enquote {\bibinfo {title} {{talk given at
  Fundamental Physics Prize Symposium, Nov. 10, 2014. and Stanford SITP
  seminars, Nov. 11 and Dec. 18, 2014.}}}\ }\href@noop {} {\  (\bibinfo {year}
  {2014})}\BibitemShut {NoStop}%
\bibitem [{\citenamefont {Kitaev}\ and\ \citenamefont
  {Suh}(2018)}]{Kitaev:2017awl}%
  \BibitemOpen
  \bibfield  {author} {\bibinfo {author} {\bibfnamefont {Alexei}\ \bibnamefont
  {Kitaev}}\ and\ \bibinfo {author} {\bibfnamefont {S.~Josephine}\ \bibnamefont
  {Suh}},\ }\bibfield  {title} {\enquote {\bibinfo {title} {{The soft mode in
  the Sachdev-Ye-Kitaev model and its gravity dual}},}\ }\href {\doibase
  10.1007/JHEP05(2018)183} {\bibfield  {journal} {\bibinfo  {journal} {JHEP}\
  }\textbf {\bibinfo {volume} {05}},\ \bibinfo {pages} {183} (\bibinfo {year}
  {2018})},\ \Eprint {http://arxiv.org/abs/1711.08467} {arXiv:1711.08467
  [hep-th]} \BibitemShut {NoStop}%
%%CITATION = ARXIV:1711.08467;%%
\bibitem [{\citenamefont {Polchinski}(2015)}]{Polchinski:2015cea}%
  \BibitemOpen
  \bibfield  {author} {\bibinfo {author} {\bibfnamefont {Joseph}\ \bibnamefont
  {Polchinski}},\ }\bibfield  {title} {\enquote {\bibinfo {title} {{Chaos in
  the black hole S-matrix}},}\ }\href@noop {} {\  (\bibinfo {year} {2015})},\
  \Eprint {http://arxiv.org/abs/1505.08108} {arXiv:1505.08108 [hep-th]}
  \BibitemShut {NoStop}%
%%CITATION = ARXIV:1505.08108;%%
\bibitem [{\citenamefont {Polchinski}\ and\ \citenamefont
  {Rosenhaus}(2016)}]{Polchinski:2016xgd}%
  \BibitemOpen
  \bibfield  {author} {\bibinfo {author} {\bibfnamefont {Joseph}\ \bibnamefont
  {Polchinski}}\ and\ \bibinfo {author} {\bibfnamefont {Vladimir}\ \bibnamefont
  {Rosenhaus}},\ }\bibfield  {title} {\enquote {\bibinfo {title} {{The Spectrum
  in the Sachdev-Ye-Kitaev Model}},}\ }\href {\doibase 10.1007/JHEP04(2016)001}
  {\bibfield  {journal} {\bibinfo  {journal} {JHEP}\ }\textbf {\bibinfo
  {volume} {04}},\ \bibinfo {pages} {001} (\bibinfo {year} {2016})},\ \Eprint
  {http://arxiv.org/abs/1601.06768} {arXiv:1601.06768 [hep-th]} \BibitemShut
  {NoStop}%
%%CITATION = ARXIV:1601.06768;%%
\bibitem [{\citenamefont {Jensen}(2016)}]{Jensen:2016pah}%
  \BibitemOpen
  \bibfield  {author} {\bibinfo {author} {\bibfnamefont {Kristan}\ \bibnamefont
  {Jensen}},\ }\bibfield  {title} {\enquote {\bibinfo {title} {{Chaos in
  AdS$_2$ Holography}},}\ }\href {\doibase 10.1103/PhysRevLett.117.111601}
  {\bibfield  {journal} {\bibinfo  {journal} {Phys. Rev. Lett.}\ }\textbf
  {\bibinfo {volume} {117}},\ \bibinfo {pages} {111601} (\bibinfo {year}
  {2016})},\ \Eprint {http://arxiv.org/abs/1605.06098} {arXiv:1605.06098
  [hep-th]} \BibitemShut {NoStop}%
%%CITATION = ARXIV:1605.06098;%%
\bibitem [{\citenamefont {Maldacena}\ and\ \citenamefont
  {Stanford}(2016)}]{Maldacena:2016hyu}%
  \BibitemOpen
  \bibfield  {author} {\bibinfo {author} {\bibfnamefont {Juan}\ \bibnamefont
  {Maldacena}}\ and\ \bibinfo {author} {\bibfnamefont {Douglas}\ \bibnamefont
  {Stanford}},\ }\bibfield  {title} {\enquote {\bibinfo {title} {{Remarks on
  the Sachdev-Ye-Kitaev model}},}\ }\href {\doibase 10.1103/PhysRevD.94.106002}
  {\bibfield  {journal} {\bibinfo  {journal} {Phys. Rev.}\ }\textbf {\bibinfo
  {volume} {D94}},\ \bibinfo {pages} {106002} (\bibinfo {year} {2016})},\
  \Eprint {http://arxiv.org/abs/1604.07818} {arXiv:1604.07818 [hep-th]}
  \BibitemShut {NoStop}%
%%CITATION = ARXIV:1604.07818;%%
\bibitem [{\citenamefont {Maldacena}\ \emph
  {et~al.}(2016{\natexlab{b}})\citenamefont {Maldacena}, \citenamefont
  {Stanford},\ and\ \citenamefont {Yang}}]{Maldacena:2016upp}%
  \BibitemOpen
  \bibfield  {author} {\bibinfo {author} {\bibfnamefont {Juan}\ \bibnamefont
  {Maldacena}}, \bibinfo {author} {\bibfnamefont {Douglas}\ \bibnamefont
  {Stanford}}, \ and\ \bibinfo {author} {\bibfnamefont {Zhenbin}\ \bibnamefont
  {Yang}},\ }\bibfield  {title} {\enquote {\bibinfo {title} {{Conformal
  symmetry and its breaking in two dimensional Nearly Anti-de-Sitter space}},}\
  }\href {\doibase 10.1093/ptep/ptw124} {\bibfield  {journal} {\bibinfo
  {journal} {PTEP}\ }\textbf {\bibinfo {volume} {2016}},\ \bibinfo {pages}
  {12C104} (\bibinfo {year} {2016}{\natexlab{b}})},\ \Eprint
  {http://arxiv.org/abs/1606.01857} {arXiv:1606.01857 [hep-th]} \BibitemShut
  {NoStop}%
%%CITATION = ARXIV:1606.01857;%%
\bibitem [{\citenamefont {Gu}\ \emph {et~al.}(2017)\citenamefont {Gu},
  \citenamefont {Qi},\ and\ \citenamefont {Stanford}}]{Gu:2016oyy}%
  \BibitemOpen
  \bibfield  {author} {\bibinfo {author} {\bibfnamefont {Yingfei}\ \bibnamefont
  {Gu}}, \bibinfo {author} {\bibfnamefont {Xiao-Liang}\ \bibnamefont {Qi}}, \
  and\ \bibinfo {author} {\bibfnamefont {Douglas}\ \bibnamefont {Stanford}},\
  }\bibfield  {title} {\enquote {\bibinfo {title} {{Local criticality,
  diffusion and chaos in generalized Sachdev-Ye-Kitaev models}},}\ }\href
  {\doibase 10.1007/JHEP05(2017)125} {\bibfield  {journal} {\bibinfo  {journal}
  {JHEP}\ }\textbf {\bibinfo {volume} {05}},\ \bibinfo {pages} {125} (\bibinfo
  {year} {2017})},\ \Eprint {http://arxiv.org/abs/1609.07832} {arXiv:1609.07832
  [hep-th]} \BibitemShut {NoStop}%
%%CITATION = ARXIV:1609.07832;%%
\bibitem [{\citenamefont {Davison}\ \emph {et~al.}(2017)\citenamefont
  {Davison}, \citenamefont {Fu}, \citenamefont {Georges}, \citenamefont {Gu},
  \citenamefont {Jensen},\ and\ \citenamefont {Sachdev}}]{Davison:2016ngz}%
  \BibitemOpen
  \bibfield  {author} {\bibinfo {author} {\bibfnamefont {Richard~A.}\
  \bibnamefont {Davison}}, \bibinfo {author} {\bibfnamefont {Wenbo}\
  \bibnamefont {Fu}}, \bibinfo {author} {\bibfnamefont {Antoine}\ \bibnamefont
  {Georges}}, \bibinfo {author} {\bibfnamefont {Yingfei}\ \bibnamefont {Gu}},
  \bibinfo {author} {\bibfnamefont {Kristan}\ \bibnamefont {Jensen}}, \ and\
  \bibinfo {author} {\bibfnamefont {Subir}\ \bibnamefont {Sachdev}},\
  }\bibfield  {title} {\enquote {\bibinfo {title} {{Thermoelectric transport in
  disordered metals without quasiparticles: The Sachdev-Ye-Kitaev models and
  holography}},}\ }\href {\doibase 10.1103/PhysRevB.95.155131} {\bibfield
  {journal} {\bibinfo  {journal} {Phys. Rev.}\ }\textbf {\bibinfo {volume}
  {B95}},\ \bibinfo {pages} {155131} (\bibinfo {year} {2017})},\ \Eprint
  {http://arxiv.org/abs/1612.00849} {arXiv:1612.00849 [cond-mat.str-el]}
  \BibitemShut {NoStop}%
%%CITATION = ARXIV:1612.00849;%%
\bibitem [{\citenamefont {Patel}\ and\ \citenamefont
  {Sachdev}(2017)}]{Patel:2016wdy}%
  \BibitemOpen
  \bibfield  {author} {\bibinfo {author} {\bibfnamefont {Aavishkar~A.}\
  \bibnamefont {Patel}}\ and\ \bibinfo {author} {\bibfnamefont {Subir}\
  \bibnamefont {Sachdev}},\ }\bibfield  {title} {\enquote {\bibinfo {title}
  {{Quantum chaos on a critical Fermi surface}},}\ }\href {\doibase
  10.1073/pnas.1618185114} {\bibfield  {journal} {\bibinfo  {journal} {Proc.
  Nat. Acad. Sci.}\ }\textbf {\bibinfo {volume} {114}},\ \bibinfo {pages}
  {1844--1849} (\bibinfo {year} {2017})},\ \Eprint
  {http://arxiv.org/abs/1611.00003} {arXiv:1611.00003 [cond-mat.str-el]}
  \BibitemShut {NoStop}%
%%CITATION = ARXIV:1611.00003;%%
\bibitem [{\citenamefont {Aleiner}\ \emph {et~al.}(2016)\citenamefont
  {Aleiner}, \citenamefont {Faoro},\ and\ \citenamefont
  {Ioffe}}]{Aleiner:2016eni}%
  \BibitemOpen
  \bibfield  {author} {\bibinfo {author} {\bibfnamefont {Igor~L.}\ \bibnamefont
  {Aleiner}}, \bibinfo {author} {\bibfnamefont {Lara}\ \bibnamefont {Faoro}}, \
  and\ \bibinfo {author} {\bibfnamefont {Lev~B.}\ \bibnamefont {Ioffe}},\
  }\bibfield  {title} {\enquote {\bibinfo {title} {{Microscopic model of
  quantum butterfly effect: out-of-time-order correlators and traveling
  combustion waves}},}\ }\href {\doibase 10.1016/j.aop.2016.09.006} {\bibfield
  {journal} {\bibinfo  {journal} {Annals Phys.}\ }\textbf {\bibinfo {volume}
  {375}},\ \bibinfo {pages} {378--406} (\bibinfo {year} {2016})},\ \Eprint
  {http://arxiv.org/abs/1609.01251} {arXiv:1609.01251 [cond-mat.stat-mech]}
  \BibitemShut {NoStop}%
%%CITATION = ARXIV:1609.01251;%%
\bibitem [{\citenamefont {Nahum}\ \emph {et~al.}(2018)\citenamefont {Nahum},
  \citenamefont {Vijay},\ and\ \citenamefont {Haah}}]{Nahum:2017yvy}%
  \BibitemOpen
  \bibfield  {author} {\bibinfo {author} {\bibfnamefont {Adam}\ \bibnamefont
  {Nahum}}, \bibinfo {author} {\bibfnamefont {Sagar}\ \bibnamefont {Vijay}}, \
  and\ \bibinfo {author} {\bibfnamefont {Jeongwan}\ \bibnamefont {Haah}},\
  }\bibfield  {title} {\enquote {\bibinfo {title} {{Operator Spreading in
  Random Unitary Circuits}},}\ }\href {\doibase 10.1103/PhysRevX.8.021014}
  {\bibfield  {journal} {\bibinfo  {journal} {Phys. Rev.}\ }\textbf {\bibinfo
  {volume} {X8}},\ \bibinfo {pages} {021014} (\bibinfo {year} {2018})},\
  \Eprint {http://arxiv.org/abs/1705.08975} {arXiv:1705.08975
  [cond-mat.str-el]} \BibitemShut {NoStop}%
%%CITATION = ARXIV:1705.08975;%%
\bibitem [{\citenamefont {Khemani}\ \emph {et~al.}(2018)\citenamefont
  {Khemani}, \citenamefont {Huse},\ and\ \citenamefont
  {Nahum}}]{Khemani:2018sdn}%
  \BibitemOpen
  \bibfield  {author} {\bibinfo {author} {\bibfnamefont {Vedika}\ \bibnamefont
  {Khemani}}, \bibinfo {author} {\bibfnamefont {David~A.}\ \bibnamefont
  {Huse}}, \ and\ \bibinfo {author} {\bibfnamefont {Adam}\ \bibnamefont
  {Nahum}},\ }\bibfield  {title} {\enquote {\bibinfo {title}
  {{Velocity-dependent Lyapunov exponents in many-body quantum, semi-classical,
  and classical chaos}},}\ }\href@noop {} {\  (\bibinfo {year} {2018})},\
  \Eprint {http://arxiv.org/abs/1803.05902} {arXiv:1803.05902
  [cond-mat.stat-mech]} \BibitemShut {NoStop}%
%%CITATION = ARXIV:1803.05902;%%
\bibitem [{\citenamefont {Kukuljan}\ \emph {et~al.}(2017)\citenamefont
  {Kukuljan}, \citenamefont {Grozdanov},\ and\ \citenamefont
  {Prosen}}]{Kukuljan:2017xag}%
  \BibitemOpen
  \bibfield  {author} {\bibinfo {author} {\bibfnamefont {Ivan}\ \bibnamefont
  {Kukuljan}}, \bibinfo {author} {\bibfnamefont {Saso}\ \bibnamefont
  {Grozdanov}}, \ and\ \bibinfo {author} {\bibfnamefont {Tomaz}\ \bibnamefont
  {Prosen}},\ }\bibfield  {title} {\enquote {\bibinfo {title} {{Weak Quantum
  Chaos}},}\ }\href {\doibase 10.1103/PhysRevB.96.060301} {\bibfield  {journal}
  {\bibinfo  {journal} {Phys. Rev.}\ }\textbf {\bibinfo {volume} {B96}},\
  \bibinfo {pages} {060301} (\bibinfo {year} {2017})},\ \Eprint
  {http://arxiv.org/abs/1701.09147} {arXiv:1701.09147 [cond-mat.stat-mech]}
  \BibitemShut {NoStop}%
%%CITATION = ARXIV:1701.09147;%%
\bibitem [{\citenamefont {Ott}(2002)}]{ott_2002}%
  \BibitemOpen
  \bibfield  {author} {\bibinfo {author} {\bibfnamefont {Edward}\ \bibnamefont
  {Ott}},\ }\href {\doibase 10.1017/CBO9780511803260} {\emph {\bibinfo {title}
  {Chaos in Dynamical Systems}}},\ \bibinfo {edition} {2nd}\ ed.\ (\bibinfo
  {publisher} {Cambridge University Press},\ \bibinfo {year}
  {2002})\BibitemShut {NoStop}%
\bibitem [{\citenamefont {Gaspard}(1998)}]{gaspard_1998}%
  \BibitemOpen
  \bibfield  {author} {\bibinfo {author} {\bibfnamefont {Pierre}\ \bibnamefont
  {Gaspard}},\ }\href {\doibase 10.1017/CBO9780511628856} {\emph {\bibinfo
  {title} {Chaos, Scattering and Statistical Mechanics}}},\ Cambridge Nonlinear
  Science Series\ (\bibinfo  {publisher} {Cambridge University Press},\
  \bibinfo {year} {1998})\BibitemShut {NoStop}%
\bibitem [{\citenamefont {Blake}(2016{\natexlab{a}})}]{Blake:2016wvh}%
  \BibitemOpen
  \bibfield  {author} {\bibinfo {author} {\bibfnamefont {Mike}\ \bibnamefont
  {Blake}},\ }\bibfield  {title} {\enquote {\bibinfo {title} {{Universal Charge
  Diffusion and the Butterfly Effect in Holographic Theories}},}\ }\href
  {\doibase 10.1103/PhysRevLett.117.091601} {\bibfield  {journal} {\bibinfo
  {journal} {Phys. Rev. Lett.}\ }\textbf {\bibinfo {volume} {117}},\ \bibinfo
  {pages} {091601} (\bibinfo {year} {2016}{\natexlab{a}})},\ \Eprint
  {http://arxiv.org/abs/1603.08510} {arXiv:1603.08510 [hep-th]} \BibitemShut
  {NoStop}%
%%CITATION = ARXIV:1603.08510;%%
\bibitem [{\citenamefont {Blake}(2016{\natexlab{b}})}]{Blake:2016sud}%
  \BibitemOpen
  \bibfield  {author} {\bibinfo {author} {\bibfnamefont {Mike}\ \bibnamefont
  {Blake}},\ }\bibfield  {title} {\enquote {\bibinfo {title} {{Universal
  Diffusion in Incoherent Black Holes}},}\ }\href {\doibase
  10.1103/PhysRevD.94.086014} {\bibfield  {journal} {\bibinfo  {journal} {Phys.
  Rev.}\ }\textbf {\bibinfo {volume} {D94}},\ \bibinfo {pages} {086014}
  (\bibinfo {year} {2016}{\natexlab{b}})},\ \Eprint
  {http://arxiv.org/abs/1604.01754} {arXiv:1604.01754 [hep-th]} \BibitemShut
  {NoStop}%
%%CITATION = ARXIV:1604.01754;%%
\bibitem [{\citenamefont {Blake}\ and\ \citenamefont
  {Donos}(2017)}]{Blake:2016jnn}%
  \BibitemOpen
  \bibfield  {author} {\bibinfo {author} {\bibfnamefont {Mike}\ \bibnamefont
  {Blake}}\ and\ \bibinfo {author} {\bibfnamefont {Aristomenis}\ \bibnamefont
  {Donos}},\ }\bibfield  {title} {\enquote {\bibinfo {title} {{Diffusion and
  Chaos from near AdS$_2$ horizons}},}\ }\href {\doibase
  10.1007/JHEP02(2017)013} {\bibfield  {journal} {\bibinfo  {journal} {JHEP}\
  }\textbf {\bibinfo {volume} {02}},\ \bibinfo {pages} {013} (\bibinfo {year}
  {2017})},\ \Eprint {http://arxiv.org/abs/1611.09380} {arXiv:1611.09380
  [hep-th]} \BibitemShut {NoStop}%
%%CITATION = ARXIV:1611.09380;%%
\bibitem [{\citenamefont {Blake}\ \emph
  {et~al.}(2017{\natexlab{a}})\citenamefont {Blake}, \citenamefont {Davison},\
  and\ \citenamefont {Sachdev}}]{Blake:2017qgd}%
  \BibitemOpen
  \bibfield  {author} {\bibinfo {author} {\bibfnamefont {Mike}\ \bibnamefont
  {Blake}}, \bibinfo {author} {\bibfnamefont {Richard~A.}\ \bibnamefont
  {Davison}}, \ and\ \bibinfo {author} {\bibfnamefont {Subir}\ \bibnamefont
  {Sachdev}},\ }\bibfield  {title} {\enquote {\bibinfo {title} {{Thermal
  diffusivity and chaos in metals without quasiparticles}},}\ }\href {\doibase
  10.1103/PhysRevD.96.106008} {\bibfield  {journal} {\bibinfo  {journal} {Phys.
  Rev.}\ }\textbf {\bibinfo {volume} {D96}},\ \bibinfo {pages} {106008}
  (\bibinfo {year} {2017}{\natexlab{a}})},\ \Eprint
  {http://arxiv.org/abs/1705.07896} {arXiv:1705.07896 [hep-th]} \BibitemShut
  {NoStop}%
%%CITATION = ARXIV:1705.07896;%%
\bibitem [{\citenamefont {Grozdanov}\ \emph
  {et~al.}(2018{\natexlab{a}})\citenamefont {Grozdanov}, \citenamefont
  {Schalm},\ and\ \citenamefont {Scopelliti}}]{Grozdanov:2017ajz}%
  \BibitemOpen
  \bibfield  {author} {\bibinfo {author} {\bibfnamefont {Saso}\ \bibnamefont
  {Grozdanov}}, \bibinfo {author} {\bibfnamefont {Koenraad}\ \bibnamefont
  {Schalm}}, \ and\ \bibinfo {author} {\bibfnamefont {Vincenzo}\ \bibnamefont
  {Scopelliti}},\ }\bibfield  {title} {\enquote {\bibinfo {title} {{Black hole
  scrambling from hydrodynamics}},}\ }\href {\doibase
  10.1103/PhysRevLett.120.231601} {\bibfield  {journal} {\bibinfo  {journal}
  {Phys. Rev. Lett.}\ }\textbf {\bibinfo {volume} {120}},\ \bibinfo {pages}
  {231601} (\bibinfo {year} {2018}{\natexlab{a}})},\ \Eprint
  {http://arxiv.org/abs/1710.00921} {arXiv:1710.00921 [hep-th]} \BibitemShut
  {NoStop}%
%%CITATION = ARXIV:1710.00921;%%
\bibitem [{\citenamefont {Blake}\ \emph
  {et~al.}(2017{\natexlab{b}})\citenamefont {Blake}, \citenamefont {Lee},\ and\
  \citenamefont {Liu}}]{Blake:2017ris}%
  \BibitemOpen
  \bibfield  {author} {\bibinfo {author} {\bibfnamefont {Mike}\ \bibnamefont
  {Blake}}, \bibinfo {author} {\bibfnamefont {Hyunseok}\ \bibnamefont {Lee}}, \
  and\ \bibinfo {author} {\bibfnamefont {Hong}\ \bibnamefont {Liu}},\
  }\bibfield  {title} {\enquote {\bibinfo {title} {{A quantum hydrodynamical
  description for scrambling and many-body chaos}},}\ }\href@noop {} {\
  (\bibinfo {year} {2017}{\natexlab{b}})},\ \Eprint
  {http://arxiv.org/abs/1801.00010} {arXiv:1801.00010 [hep-th]} \BibitemShut
  {NoStop}%
%%CITATION = ARXIV:1801.00010;%%
\bibitem [{\citenamefont {Grozdanov}\ \emph
  {et~al.}(2018{\natexlab{b}})\citenamefont {Grozdanov}, \citenamefont
  {Schalm},\ and\ \citenamefont {Scopelliti}}]{Grozdanov:2018atb}%
  \BibitemOpen
  \bibfield  {author} {\bibinfo {author} {\bibfnamefont {Saso}\ \bibnamefont
  {Grozdanov}}, \bibinfo {author} {\bibfnamefont {Koenraad}\ \bibnamefont
  {Schalm}}, \ and\ \bibinfo {author} {\bibfnamefont {Vincenzo}\ \bibnamefont
  {Scopelliti}},\ }\bibfield  {title} {\enquote {\bibinfo {title} {{Kinetic
  theory for classical and quantum many-body chaos}},}\ }\href@noop {} {\
  (\bibinfo {year} {2018}{\natexlab{b}})},\ \Eprint
  {http://arxiv.org/abs/1804.09182} {arXiv:1804.09182 [hep-th]} \BibitemShut
  {NoStop}%
%%CITATION = ARXIV:1804.09182;%%
\bibitem [{\citenamefont {Lucas}(2017)}]{Lucas:2017ibu}%
  \BibitemOpen
  \bibfield  {author} {\bibinfo {author} {\bibfnamefont {Andrew}\ \bibnamefont
  {Lucas}},\ }\bibfield  {title} {\enquote {\bibinfo {title} {{Constraints on
  hydrodynamics from many-body quantum chaos}},}\ }\href@noop {} {\  (\bibinfo
  {year} {2017})},\ \Eprint {http://arxiv.org/abs/1710.01005} {arXiv:1710.01005
  [hep-th]} \BibitemShut {NoStop}%
%%CITATION = ARXIV:1710.01005;%%
\bibitem [{\citenamefont {Hartman}\ \emph {et~al.}(2017)\citenamefont
  {Hartman}, \citenamefont {Hartnoll},\ and\ \citenamefont
  {Mahajan}}]{Hartman:2017hhp}%
  \BibitemOpen
  \bibfield  {author} {\bibinfo {author} {\bibfnamefont {Thomas}\ \bibnamefont
  {Hartman}}, \bibinfo {author} {\bibfnamefont {Sean~A.}\ \bibnamefont
  {Hartnoll}}, \ and\ \bibinfo {author} {\bibfnamefont {Raghu}\ \bibnamefont
  {Mahajan}},\ }\bibfield  {title} {\enquote {\bibinfo {title} {{Upper Bound on
  Diffusivity}},}\ }\href {\doibase 10.1103/PhysRevLett.119.141601} {\bibfield
  {journal} {\bibinfo  {journal} {Phys. Rev. Lett.}\ }\textbf {\bibinfo
  {volume} {119}},\ \bibinfo {pages} {141601} (\bibinfo {year} {2017})},\
  \Eprint {http://arxiv.org/abs/1706.00019} {arXiv:1706.00019 [hep-th]}
  \BibitemShut {NoStop}%
%%CITATION = ARXIV:1706.00019;%%
\bibitem [{\citenamefont {Davison}\ \emph {et~al.}(2018)\citenamefont
  {Davison}, \citenamefont {Gentle},\ and\ \citenamefont
  {Gouteraux}}]{Davison:2018ofp}%
  \BibitemOpen
  \bibfield  {author} {\bibinfo {author} {\bibfnamefont {Richard~A.}\
  \bibnamefont {Davison}}, \bibinfo {author} {\bibfnamefont {Simon~A.}\
  \bibnamefont {Gentle}}, \ and\ \bibinfo {author} {\bibfnamefont {Blaise}\
  \bibnamefont {Gouteraux}},\ }\bibfield  {title} {\enquote {\bibinfo {title}
  {{Slow relaxation and diffusion in holographic quantum critical phases}},}\
  }\href@noop {} {\  (\bibinfo {year} {2018})},\ \Eprint
  {http://arxiv.org/abs/1808.05659} {arXiv:1808.05659 [hep-th]} \BibitemShut
  {NoStop}%
%%CITATION = ARXIV:1808.05659;%%
\bibitem [{\citenamefont {Haehl}\ and\ \citenamefont {Rozali}(2018)}]{moshe}%
  \BibitemOpen
  \bibfield  {author} {\bibinfo {author} {\bibfnamefont {Felix~M.}\
  \bibnamefont {Haehl}}\ and\ \bibinfo {author} {\bibfnamefont {Moshe}\
  \bibnamefont {Rozali}},\ }\bibfield  {title} {\enquote {\bibinfo {title}
  {{Effective Field Theory for Chaotic CFTs}},}\ }\href@noop {} {\  (\bibinfo
  {year} {2018})},\ \Eprint {http://arxiv.org/abs/1808.02898} {arXiv:1808.02898
  [hep-th]} \BibitemShut {NoStop}%
%%CITATION = ARXIV:1808.02898;%%
\bibitem [{\citenamefont {Andrade}\ and\ \citenamefont
  {Withers}(2014)}]{Andrade:2013gsa}%
  \BibitemOpen
  \bibfield  {author} {\bibinfo {author} {\bibfnamefont {Tomas}\ \bibnamefont
  {Andrade}}\ and\ \bibinfo {author} {\bibfnamefont {Benjamin}\ \bibnamefont
  {Withers}},\ }\bibfield  {title} {\enquote {\bibinfo {title} {{A simple
  holographic model of momentum relaxation}},}\ }\href {\doibase
  10.1007/JHEP05(2014)101} {\bibfield  {journal} {\bibinfo  {journal} {JHEP}\
  }\textbf {\bibinfo {volume} {05}},\ \bibinfo {pages} {101} (\bibinfo {year}
  {2014})},\ \Eprint {http://arxiv.org/abs/1311.5157} {arXiv:1311.5157
  [hep-th]} \BibitemShut {NoStop}%
%%CITATION = ARXIV:1311.5157;%%
\bibitem [{\citenamefont {Davison}\ and\ \citenamefont
  {Gouteraux}(2015)}]{Davison:2014lua}%
  \BibitemOpen
  \bibfield  {author} {\bibinfo {author} {\bibfnamefont {Richard~A.}\
  \bibnamefont {Davison}}\ and\ \bibinfo {author} {\bibfnamefont {Blaise}\
  \bibnamefont {Gouteraux}},\ }\bibfield  {title} {\enquote {\bibinfo {title}
  {{Momentum dissipation and effective theories of coherent and incoherent
  transport}},}\ }\href {\doibase 10.1007/JHEP01(2015)039} {\bibfield
  {journal} {\bibinfo  {journal} {JHEP}\ }\textbf {\bibinfo {volume} {01}},\
  \bibinfo {pages} {039} (\bibinfo {year} {2015})},\ \Eprint
  {http://arxiv.org/abs/1411.1062} {arXiv:1411.1062 [hep-th]} \BibitemShut
  {NoStop}%
%%CITATION = ARXIV:1411.1062;%%
\bibitem [{\citenamefont {Roberts}\ and\ \citenamefont
  {Swingle}(2016)}]{Roberts:2016wdl}%
  \BibitemOpen
  \bibfield  {author} {\bibinfo {author} {\bibfnamefont {Daniel~A.}\
  \bibnamefont {Roberts}}\ and\ \bibinfo {author} {\bibfnamefont {Brian}\
  \bibnamefont {Swingle}},\ }\bibfield  {title} {\enquote {\bibinfo {title}
  {{Lieb-Robinson Bound and the Butterfly Effect in Quantum Field Theories}},}\
  }\href {\doibase 10.1103/PhysRevLett.117.091602} {\bibfield  {journal}
  {\bibinfo  {journal} {Phys. Rev. Lett.}\ }\textbf {\bibinfo {volume} {117}},\
  \bibinfo {pages} {091602} (\bibinfo {year} {2016})},\ \Eprint
  {http://arxiv.org/abs/1603.09298} {arXiv:1603.09298 [hep-th]} \BibitemShut
  {NoStop}%
%%CITATION = ARXIV:1603.09298;%%
\bibitem [{\citenamefont {Son}\ and\ \citenamefont
  {Starinets}(2002)}]{Son:2002sd}%
  \BibitemOpen
  \bibfield  {author} {\bibinfo {author} {\bibfnamefont {Dam~T.}\ \bibnamefont
  {Son}}\ and\ \bibinfo {author} {\bibfnamefont {Andrei~O.}\ \bibnamefont
  {Starinets}},\ }\bibfield  {title} {\enquote {\bibinfo {title} {{Minkowski
  space correlators in AdS / CFT correspondence: Recipe and applications}},}\
  }\href {\doibase 10.1088/1126-6708/2002/09/042} {\bibfield  {journal}
  {\bibinfo  {journal} {JHEP}\ }\textbf {\bibinfo {volume} {09}},\ \bibinfo
  {pages} {042} (\bibinfo {year} {2002})},\ \Eprint
  {http://arxiv.org/abs/hep-th/0205051} {arXiv:hep-th/0205051 [hep-th]}
  \BibitemShut {NoStop}%
%%CITATION = HEP-TH/0205051;%%
\bibitem [{\citenamefont {Kovtun}\ and\ \citenamefont
  {Starinets}(2005)}]{Kovtun:2005ev}%
  \BibitemOpen
  \bibfield  {author} {\bibinfo {author} {\bibfnamefont {Pavel~K.}\
  \bibnamefont {Kovtun}}\ and\ \bibinfo {author} {\bibfnamefont {Andrei~O.}\
  \bibnamefont {Starinets}},\ }\bibfield  {title} {\enquote {\bibinfo {title}
  {{Quasinormal modes and holography}},}\ }\href {\doibase
  10.1103/PhysRevD.72.086009} {\bibfield  {journal} {\bibinfo  {journal} {Phys.
  Rev.}\ }\textbf {\bibinfo {volume} {D72}},\ \bibinfo {pages} {086009}
  (\bibinfo {year} {2005})},\ \Eprint {http://arxiv.org/abs/hep-th/0506184}
  {arXiv:hep-th/0506184 [hep-th]} \BibitemShut {NoStop}%
%%CITATION = HEP-TH/0506184;%%
\bibitem [{\citenamefont {Herzog}(2002)}]{Herzog:2002fn}%
  \BibitemOpen
  \bibfield  {author} {\bibinfo {author} {\bibfnamefont {Christopher~P.}\
  \bibnamefont {Herzog}},\ }\bibfield  {title} {\enquote {\bibinfo {title}
  {{The Hydrodynamics of M theory}},}\ }\href {\doibase
  10.1088/1126-6708/2002/12/026} {\bibfield  {journal} {\bibinfo  {journal}
  {JHEP}\ }\textbf {\bibinfo {volume} {12}},\ \bibinfo {pages} {026} (\bibinfo
  {year} {2002})},\ \Eprint {http://arxiv.org/abs/hep-th/0210126}
  {arXiv:hep-th/0210126 [hep-th]} \BibitemShut {NoStop}%
%%CITATION = HEP-TH/0210126;%%
\bibitem [{\citenamefont {Grozdanov}\ and\ \citenamefont
  {Kaplis}(2016)}]{Grozdanov:2015kqa}%
  \BibitemOpen
  \bibfield  {author} {\bibinfo {author} {\bibfnamefont {Saso}\ \bibnamefont
  {Grozdanov}}\ and\ \bibinfo {author} {\bibfnamefont {Nikolaos}\ \bibnamefont
  {Kaplis}},\ }\bibfield  {title} {\enquote {\bibinfo {title} {{Constructing
  higher-order hydrodynamics: The third order}},}\ }\href {\doibase
  10.1103/PhysRevD.93.066012} {\bibfield  {journal} {\bibinfo  {journal} {Phys.
  Rev.}\ }\textbf {\bibinfo {volume} {D93}},\ \bibinfo {pages} {066012}
  (\bibinfo {year} {2016})},\ \Eprint {http://arxiv.org/abs/1507.02461}
  {arXiv:1507.02461 [hep-th]} \BibitemShut {NoStop}%
%%CITATION = ARXIV:1507.02461;%%
\bibitem [{\citenamefont {Davison}\ and\ \citenamefont
  {Parnachev}(2013)}]{Davison:2013bxa}%
  \BibitemOpen
  \bibfield  {author} {\bibinfo {author} {\bibfnamefont {Richard~A.}\
  \bibnamefont {Davison}}\ and\ \bibinfo {author} {\bibfnamefont {Andrei}\
  \bibnamefont {Parnachev}},\ }\bibfield  {title} {\enquote {\bibinfo {title}
  {{Hydrodynamics of cold holographic matter}},}\ }\href {\doibase
  10.1007/JHEP06(2013)100} {\bibfield  {journal} {\bibinfo  {journal} {JHEP}\
  }\textbf {\bibinfo {volume} {06}},\ \bibinfo {pages} {100} (\bibinfo {year}
  {2013})},\ \Eprint {http://arxiv.org/abs/1303.6334} {arXiv:1303.6334
  [hep-th]} \BibitemShut {NoStop}%
%%CITATION = ARXIV:1303.6334;%%
\bibitem [{\citenamefont {Compere}(2012)}]{Compere:2012jk}%
  \BibitemOpen
  \bibfield  {author} {\bibinfo {author} {\bibfnamefont {Geoffrey}\
  \bibnamefont {Compere}},\ }\bibfield  {title} {\enquote {\bibinfo {title}
  {{The Kerr/CFT correspondence and its extensions}},}\ }\href {\doibase
  10.1007/s41114-017-0003-2} {\bibfield  {journal} {\bibinfo  {journal} {Living
  Rev. Rel.}\ }\textbf {\bibinfo {volume} {15}},\ \bibinfo {pages} {11}
  (\bibinfo {year} {2012})},\ \bibinfo {note} {[Living Rev.
  Rel.20,no.1,1(2017)]},\ \Eprint {http://arxiv.org/abs/1203.3561}
  {arXiv:1203.3561 [hep-th]} \BibitemShut {NoStop}%
%%CITATION = ARXIV:1203.3561;%%
\bibitem [{\citenamefont {Grozdanov}\ and\ \citenamefont
  {Starinets}(2017)}]{Grozdanov:2016fkt}%
  \BibitemOpen
  \bibfield  {author} {\bibinfo {author} {\bibfnamefont {Saso}\ \bibnamefont
  {Grozdanov}}\ and\ \bibinfo {author} {\bibfnamefont {Andrei~O.}\ \bibnamefont
  {Starinets}},\ }\bibfield  {title} {\enquote {\bibinfo {title} {{Second-order
  transport, quasinormal modes and zero-viscosity limit in the Gauss-Bonnet
  holographic fluid}},}\ }\href {\doibase 10.1007/JHEP03(2017)166} {\bibfield
  {journal} {\bibinfo  {journal} {JHEP}\ }\textbf {\bibinfo {volume} {03}},\
  \bibinfo {pages} {166} (\bibinfo {year} {2017})},\ \Eprint
  {http://arxiv.org/abs/1611.07053} {arXiv:1611.07053 [hep-th]} \BibitemShut
  {NoStop}%
%%CITATION = ARXIV:1611.07053;%%
\bibitem [{\citenamefont {Bueno}\ \emph {et~al.}(2018)\citenamefont {Bueno},
  \citenamefont {Cano},\ and\ \citenamefont {Ruiperez}}]{Bueno:2018xqc}%
  \BibitemOpen
  \bibfield  {author} {\bibinfo {author} {\bibfnamefont {Pablo}\ \bibnamefont
  {Bueno}}, \bibinfo {author} {\bibfnamefont {Pablo~A.}\ \bibnamefont {Cano}},
  \ and\ \bibinfo {author} {\bibfnamefont {Alejandro}\ \bibnamefont
  {Ruiperez}},\ }\bibfield  {title} {\enquote {\bibinfo {title} {{Holographic
  studies of Einsteinian cubic gravity}},}\ }\href {\doibase
  10.1007/JHEP03(2018)150} {\bibfield  {journal} {\bibinfo  {journal} {JHEP}\
  }\textbf {\bibinfo {volume} {03}},\ \bibinfo {pages} {150} (\bibinfo {year}
  {2018})},\ \Eprint {http://arxiv.org/abs/1802.00018} {arXiv:1802.00018
  [hep-th]} \BibitemShut {NoStop}%
%%CITATION = ARXIV:1802.00018;%%
\bibitem [{\citenamefont {Donos}\ \emph {et~al.}(2018)\citenamefont {Donos},
  \citenamefont {Gauntlett},\ and\ \citenamefont {Ziogas}}]{Donos:2017ihe}%
  \BibitemOpen
  \bibfield  {author} {\bibinfo {author} {\bibfnamefont {Aristomenis}\
  \bibnamefont {Donos}}, \bibinfo {author} {\bibfnamefont {Jerome~P.}\
  \bibnamefont {Gauntlett}}, \ and\ \bibinfo {author} {\bibfnamefont {Vaios}\
  \bibnamefont {Ziogas}},\ }\bibfield  {title} {\enquote {\bibinfo {title}
  {{Diffusion for Holographic Lattices}},}\ }\href {\doibase
  10.1007/JHEP03(2018)056} {\bibfield  {journal} {\bibinfo  {journal} {JHEP}\
  }\textbf {\bibinfo {volume} {03}},\ \bibinfo {pages} {056} (\bibinfo {year}
  {2018})},\ \Eprint {http://arxiv.org/abs/1710.04221} {arXiv:1710.04221
  [hep-th]} \BibitemShut {NoStop}%
%%CITATION = ARXIV:1710.04221;%%
\bibitem [{\citenamefont {Brigante}\ \emph {et~al.}(2008)\citenamefont
  {Brigante}, \citenamefont {Liu}, \citenamefont {Myers}, \citenamefont
  {Shenker},\ and\ \citenamefont {Yaida}}]{Brigante:2007nu}%
  \BibitemOpen
  \bibfield  {author} {\bibinfo {author} {\bibfnamefont {Mauro}\ \bibnamefont
  {Brigante}}, \bibinfo {author} {\bibfnamefont {Hong}\ \bibnamefont {Liu}},
  \bibinfo {author} {\bibfnamefont {Robert~C.}\ \bibnamefont {Myers}}, \bibinfo
  {author} {\bibfnamefont {Stephen}\ \bibnamefont {Shenker}}, \ and\ \bibinfo
  {author} {\bibfnamefont {Sho}\ \bibnamefont {Yaida}},\ }\bibfield  {title}
  {\enquote {\bibinfo {title} {{Viscosity Bound Violation in Higher Derivative
  Gravity}},}\ }\href {\doibase 10.1103/PhysRevD.77.126006} {\bibfield
  {journal} {\bibinfo  {journal} {Phys. Rev.}\ }\textbf {\bibinfo {volume}
  {D77}},\ \bibinfo {pages} {126006} (\bibinfo {year} {2008})},\ \Eprint
  {http://arxiv.org/abs/0712.0805} {arXiv:0712.0805 [hep-th]} \BibitemShut
  {NoStop}%
%%CITATION = ARXIV:0712.0805;%%
\bibitem [{\citenamefont {Grozdanov}\ \emph {et~al.}(2016)\citenamefont
  {Grozdanov}, \citenamefont {Kaplis},\ and\ \citenamefont
  {Starinets}}]{Grozdanov:2016vgg}%
  \BibitemOpen
  \bibfield  {author} {\bibinfo {author} {\bibfnamefont {Saso}\ \bibnamefont
  {Grozdanov}}, \bibinfo {author} {\bibfnamefont {Nikolaos}\ \bibnamefont
  {Kaplis}}, \ and\ \bibinfo {author} {\bibfnamefont {Andrei~O.}\ \bibnamefont
  {Starinets}},\ }\bibfield  {title} {\enquote {\bibinfo {title} {{From strong
  to weak coupling in holographic models of thermalization}},}\ }\href
  {\doibase 10.1007/JHEP07(2016)151} {\bibfield  {journal} {\bibinfo  {journal}
  {JHEP}\ }\textbf {\bibinfo {volume} {07}},\ \bibinfo {pages} {151} (\bibinfo
  {year} {2016})},\ \Eprint {http://arxiv.org/abs/1605.02173} {arXiv:1605.02173
  [hep-th]} \BibitemShut {NoStop}%
%%CITATION = ARXIV:1605.02173;%%
\bibitem [{\citenamefont {Vegh}(2013)}]{Vegh:2013sk}%
  \BibitemOpen
  \bibfield  {author} {\bibinfo {author} {\bibfnamefont {David}\ \bibnamefont
  {Vegh}},\ }\bibfield  {title} {\enquote {\bibinfo {title} {{Holography
  without translational symmetry}},}\ }\href@noop {} {\  (\bibinfo {year}
  {2013})},\ \Eprint {http://arxiv.org/abs/1301.0537} {arXiv:1301.0537
  [hep-th]} \BibitemShut {NoStop}%
%%CITATION = ARXIV:1301.0537;%%
\bibitem [{\citenamefont {Bianchi}\ \emph {et~al.}(2002)\citenamefont
  {Bianchi}, \citenamefont {Freedman},\ and\ \citenamefont
  {Skenderis}}]{Bianchi:2001kw}%
  \BibitemOpen
  \bibfield  {author} {\bibinfo {author} {\bibfnamefont {Massimo}\ \bibnamefont
  {Bianchi}}, \bibinfo {author} {\bibfnamefont {Daniel~Z.}\ \bibnamefont
  {Freedman}}, \ and\ \bibinfo {author} {\bibfnamefont {Kostas}\ \bibnamefont
  {Skenderis}},\ }\bibfield  {title} {\enquote {\bibinfo {title} {{Holographic
  renormalization}},}\ }\href {\doibase 10.1016/S0550-3213(02)00179-7}
  {\bibfield  {journal} {\bibinfo  {journal} {Nucl. Phys.}\ }\textbf {\bibinfo
  {volume} {B631}},\ \bibinfo {pages} {159--194} (\bibinfo {year} {2002})},\
  \Eprint {http://arxiv.org/abs/hep-th/0112119} {arXiv:hep-th/0112119 [hep-th]}
  \BibitemShut {NoStop}%
%%CITATION = HEP-TH/0112119;%%
\end{thebibliography}%
\end{document}